\documentclass[a4paper,11pt]{article}
\pdfoutput=1 % if your are submitting a pdflatex (i.e. if you have
\usepackage{jheppub} % for details on the use of the package, please
\usepackage[T1]{fontenc} % if needed

\newcommand{\tr}{\mbox{Tr}}

\title{Solving for the integrable boundary states of the ABJM spin chain from $KT$-relations}

 \author[1]{N.~Bai,\note{Corresponding author.}}
 \author{H.~Yang}
 \author{and M.~Z.~Shao}
 \affiliation{Department of Physics, Guangxi Normal University\\Guilin 541004, China}

\emailAdd{bainan@mailbox.gxnu.edu.cn}
\emailAdd{huiyang@stu.gxnu.edu.cn}
\emailAdd{mzshao@stu.gxnu.edu.cn}
\abstract{We study integrable boundary states of the alternating SU(4) spin chain arising in ABJM theory.  Starting from the $KT$-relation, we directly solve the integrability constraints for states with $n$-site translational invariance. For odd and even $n$, these constraints are reduced respectively to state
equations and operator equations for the elementary $n$-site block and the matrix $K(u)$.  We analyze chiral and achiral cases for $n\leq4$.  In the
1-site case we allow general operator-valued integrable pairs with an internal space, while in the remaining cases we focus on $c$-number solutions.}

\begin{document}
\maketitle
\flushbottom
\section{Introduction}

Integrable boundary states have recently attracted considerable attention in several areas of theoretical physics. The notion traces back to the study of integrable boundaries in two-dimensional integrable quantum field theory \cite{Ghoshal:1993tm}, while its meaning in quantum integrable models was clarified much more recently in the seminal works of Piroli, Pozsgay and Vernier \cite{Piroli:2017sei,Pozsgay:2018dzs}. It has found important applications in statistical and condensed matter physics, especially in the study of non-equilibrium dynamics and quantum quenches in integrable models \cite{Caux:2013ra,Wouters:2014,Pozsgay:2014,Mestyan:2017xyk,Piroli:2018ksf,Piroli:2018don,Rylands:2022gev,Rylands:2022naf}.

In holographic integrability, integrable boundary states appear naturally in defect versions of $\mathcal N=4$ SYM and in the corresponding $\rm{AdS_5/CFT_4}$ setup.  For the D3--D5 interface, one-point functions of single trace operators can be mapped to overlaps between Bethe eigenstates of the spin chain and special boundary states \cite{deLeeuw:2015hxa,Buhl-Mortensen:2015gfd,deLeeuw:2016umh}. Subsequently, one-point and higher-point correlation functions have been
extensively studied in a variety of defect CFTs and related holographic setups \cite{deLeeuw:2017dkd,Buhl-Mortensen:2017ind,DeLeeuw:2018cal, Komatsu:2020sup,Kristjansen:2020mhn,Holguin:2025bfe,Chalabi:2025nbg,Linardopoulos:2021rfq,Linardopoulos:2025ypq}. Related overlap constructions and integrability methods have also been developed for determinant operators, line defects, Coulomb-branch observables, and other holographic configurations \cite{Jiang:2019xdz,Jiang:2019zig,DeLeeuw:2019ohp,Kristjansen:2023ysz,Ivanovskiy:2024vel,Gombor:2024api,Kristjansen:2024map,Coronado:2025xwk,Demjaha:2025axy,Gombor:2025qvk}. In view of the central role played by exact overlap formulas in the study of defect CFT correlation functions and quench dynamics, Gombor has developed a systematic algebraic approach to deriving such formulas in a series of works \cite{Gombor:2021uxz,Gombor:2021hmj,Gombor:2023bez,Gombor:2024iix,Gombor:2025wvu}.

It is therefore natural to ask analogous questions in ABJM theory, the three-dimensional $\mathcal N=6$ superconformal Chern--Simons-matter theory dual to string theory on $\rm{AdS_4\times\mathbb{CP}^3}$ \cite{Aharony:2008ug}.  The planar scalar sector of this theory is described at two loops by an alternating $SU(4)$ spin chain, with neighboring sites transforming in the fundamental and anti-fundamental representations \cite{Minahan:2008hf,Bak:2008cp}. Integrable boundary states in this alternating chain have appeared in ABJM three-point functions, domain-wall defects, giant-graviton constructions, and Wilson-loop one-point functions\cite{Yang:2021hrl,Kristjansen:2021abc,Gombor:2022aqj,Yang:2022dlk,Jiang:2023cdm,Wu:2024uix}.  Direct constructions of integrable states in the
alternating $SU(4)$ chain have also been developed in \cite{Bai:2024qtg,Liu:2025uiu,Liu:2026abjmre}.  On the string-theory
side, related investigations have been studied for ABJM defect and D-brane observables \cite{Linardopoulos:2022wol,Yang:2021kot}.

For homogeneous spin chains, integrable boundary states were originally defined as states annihilated by all odd conserved charges \cite{Piroli:2017sei}, in close analogy with the criterion for integrable boundary states in two-dimensional integrable quantum field theory \cite{Ghoshal:1993tm}. For the ABJM spin chain, however, the alternating structure gives rise to two sets of conserved charges associated with the two subchains. It is therefore more appropriate to formulate the integrability condition directly in terms of relations between transfer matrices. Since the ABJM spin chain contains two transfer matrices, this formulation leads to two types of integrability conditions, chiral and achiral \cite{Gombor:2020kgu}: the chiral condition relates the two transfer matrices to each other at crossed spectral parameters, whereas the achiral condition relates a transfer matrix to itself at the crossed spectral parameter. Later, a more fundamental algebraic condition called the $KT$-relation was introduced and plays an important role in algebraic derivations and overlap computations \cite{Gombor:2021uxz,Gombor:2021hmj}. For the ABJM spin chain, the same alternating structure also gives rise to chiral and achiral versions of the $KT$-relation.

Integrable boundary states are usually taken to be translationally invariant. It is nevertheless also worthwhile to consider boundary states with $n$-site translational invariance, built from an elementary $n$-site block, especially the 2-site case, where the construction is closely related to solutions of boundary reflection equations. In this direction, the multi-site integrable boundary states in the ABJM spin chain were studied in \cite{Liu:2026abjmre}. There the fusion procedure for boundary integrable models \cite{Mezincescu:1991ke,Bai:2025mpw} was used to build suitable fused $K$-matrices from elementary soliton-preserving (SP) and soliton-non-preserving (SNP) reflection matrices, leading to a class of $2n$-site integrable boundary states.

The aim of this paper is to revisit the construction of integrable boundary states in the ABJM spin chain directly from the $KT$-relation. For an
$n$-site translationally invariant boundary state, the $KT$-relation can be reduced to a fundamental equation involving the elementary $n$-site block $|\phi\rangle$ and the auxiliary space matrix $K(u)$. We treat these two objects as the unknowns of this reduced constraint and refer to a solution as an integrable pair $(|\phi\rangle, K(u))$. In solving these $KT$-equations, we avoid treating $K(u)$ as a reflection matrix whenever possible. Instead, we solve for the boundary block and $K(u)$ simultaneously and then examine in which cases the resulting $K(u)$ is forced to satisfy a reflection equation. The concrete solution strategy is further divided into two cases according to the parity of $n$. For odd $n$, the $KT$-relation can be converted into a state equation satisfied by the $n$-site block, whereas for even $n$ it is converted into an operator equation. We then analyze these equations by combining several mathematical techniques, including group-theoretical arguments and rank analyses of matrix equations at special values of the spectral parameter. An advantage of this approach is that it allows for a broader class of solutions than those obtained from reflection equations or their fused counterparts; in particular, it provides a direct method for constructing odd-site translationally invariant integrable states that cannot be constructed by the reflection equation method.

The paper is organized as follows.  In section 2 we review the ABJM alternating spin chain, define chiral and achiral integrable states, and formulate the
corresponding $KT$-relations.  In sections 3, 4, 5 and 6, we study the 1-site, 2-site, 3-site and 4-site cases, respectively.  In appendices~\ref{app:SP}
and~\ref{app:SNP}, we present the SP- and SNP-type reflection matrices used in the main text.  In appendix~\ref{appx:matrix}, we give a simple matrix identity
that is used in the 4-site analysis.

\section{Integrable boundary states and $KT$-relations in the ABJM spin chain}
The ABJM spin chain is an alternating $SU(4)$ spin chain, in which neighboring sites transform in the fundamental representation $\mathbf{4}$ and the anti-fundamental representation $\bar{\mathbf{4}}$ of $SU(4)$, respectively. Accordingly, there are two types of $R$-matrices, which we denote as follows:
\begin{equation}\label{sec1:f3}
R_{ab}(u)=u\mathbb{I}+\mathbb{P},\quad \bar{R}_{ab}(u)=-(u+2)\mathbb{I}+\mathbb{Q}.
\end{equation}
Here the first $R$-matrix is used when $V_a$ and $V_b$ are representation spaces of the same type, whereas the second $\bar{R}$-matrix is used when they are representation spaces of different types. Here
\begin{equation}
\mathbb{P}=E_{ij}\otimes E_{ji},\quad \mathbb{Q}=E_{ij}\otimes E_{ij}.
\end{equation}
We consider a spin chain of even length $2N$ and introduce the following two monodromy matrices:
\begin{eqnarray}
&&T_0(u):=T_{0;1,2\cdots 2N-1,2N}(u)=R_{01}(u)\bar{R}_{02}(u)\cdots R_{0,2N-1}(u)\bar{R}_{0,2N}(u),\\
&&\bar{T}_{0'}(u):=\bar{T}_{0';1,2\cdots 2N-1,2N}(u)=\bar{R}_{0'1}(u){R}_{0'2}(u)\cdots \bar{R}_{0',2N-1}(u){R}_{0',2N}(u),
\end{eqnarray}
where $V_0$ and $V_{2j-1},j=1,2,\cdots N$ are fundamental representation spaces, while $V_{0'}$ and $V_{2j},j=1,2,\cdots N$ are anti-fundamental representation spaces. The spaces $V_0$ and $V_{0'}$ are conventionally called auxiliary spaces, whereas $V_1,V_2,\cdots, V_{2N}$ are quantum spaces. The transfer matrices are defined by
\begin{equation}
\tau(u)=\tr_0 T_0(u),\quad \bar{\tau}(u)=\tr_{0'}\bar{T}_{0'}(u).
\end{equation}
Since the $R$- and $\bar{R}$-matrices satisfy the Yang--Baxter equations, the transfer matrices form a mutually commuting family:
\begin{equation}
[\tau(u),\tau(v)]=0,\quad [\bar{\tau}(u),\bar{\tau}(v)]=0,\quad [\tau(u),\bar{\tau}(v)]=0.
\end{equation}
As usual, the conserved charges are obtained by expanding the logarithms of the transfer matrices around $u=0$:
\begin{equation}
Q_{n+1}=\frac{\partial^n}{\partial u^n}\log \tau(u)\bigg|_{u=0},\quad \bar{Q}_{n+1}=\frac{\partial^n}{\partial u^n}\log \bar{\tau}(u)\bigg|_{u=0},
\end{equation}
and the Hamiltonian of the ABJM spin chain can be written as the following combination of conserved charges:
\begin{equation}
H_{\rm{ABJM}}=Q_2+\bar{Q}_2.
\end{equation}
The quantum space of the full spin chain is denoted by $\mathcal{H}$:
\begin{equation}
\mathcal{H}\cong (\mathbb{C}^4)^{\otimes 2N}.
\end{equation}
We are interested in integrable boundary states defined on $\mathcal{H}$: $|\Psi\rangle\in\mathcal{H}$. Such states are usually characterized as states annihilated by all odd conserved charges. In addition, Bethe eigenstates with nonvanishing overlap with an integrable boundary state exhibit the pair structure of the Bethe roots. For the ABJM spin chain considered here, however, the conserved charges generated by $\tau(u)$ and $\bar{\tau}(u)$ are intertwined. It is therefore more natural to define integrable states through relations between transfer matrices. More precisely, for the ABJM spin chain, we define chiral and achiral integrable states as follows:
\begin{align}
{\mbox{chiral:}}\quad
\tau(u)|\Psi\rangle&=\bar{\tau}(-u-2)|\Psi\rangle,
\\
{\mbox{achiral:}}\quad
\tau(u)|\Psi\rangle&=\tau(-u-2)|\Psi\rangle.
\end{align}
A chiral integrable state leads to the pairing among Bethe roots of the same type, whereas an achiral integrable state leads to the pairing of different types.

A sufficient condition for constructing integrable states is provided by the following $KT$-relation:
\begin{align}\label{sec1:f1}
{\mbox{chiral:}}\quad
T_0(u)|\Psi\rangle K_0(u)
&= K_0(u) \bar{T}_0(-u-2)|\Psi\rangle ,
\\ \label{sec1:f2}
{\mbox{achiral:}}\quad
T_0(u)|\Psi\rangle K_0(u)
&= K_0(u) T_0(-u-2)|\Psi\rangle .
\end{align}
The $KT$-relation was introduced in the work of Gombor and Pozsgay on exact overlap formulas for integrable boundary states \cite{Gombor:2021uxz}.  The equations above are the ABJM counterpart of this relation, adapted to the alternating fundamental/anti-fundamental spin chain.  The chiral $KT$-relation implies that $K(u)$ satisfies the following ``self-consistency condition'':
\begin{equation}
\begin{split}
R_{12}(u-v)K_1(-u-1)\bar{R}_{12}(u+v)K_2(-v-1)=\\
K_2(-v-1)\bar{R}_{12}(u+v)K_1(-u-1)R_{12}(u-v),
\end{split}
\end{equation}
which indicates that $K(u)$ may be chosen as a solution of the SNP-type reflection equation. On the other hand, the achiral $KT$-relation gives
\begin{equation}
\begin{split}
R_{12}(u-v)K_1(-u-1){R}_{12}(u+v)K_2(-v-1)=\\
K_2(-v-1){R}_{12}(u+v)K_1(-u-1)R_{12}(u-v),
\end{split}
\end{equation}
which indicates that $K(u)$ may be chosen as a solution of the SP-type reflection equation. The solutions of these two reflection equations are given in appendices~\ref{app:SP} and~\ref{app:SNP}. We emphasize that these two self-consistency conditions are sufficient, but not necessary, conditions for $K(u)$. Once the form of the $R$-matrix is fixed, the $KT$-relation yields a mixed state-operator constraint equation involving both the integrable state $|\Psi\rangle$ and the matrix $K(u)$. A pair $(|\Psi\rangle, K(u))$ satisfying this constraint equation may be called an integrable pair, and such a pair gives rise to the corresponding integrability condition. 

In what follows, based on the $KT$-relation, we search for possible chiral and achiral integrable states with $n$-site ($n\leq 4$) translational invariance.

\section{Integrable 1-site states}
We first consider fully translationally invariant integrable states, which we refer to as integrable 1-site states.  Such a state can be written as
\begin{equation}
|\Psi\rangle=\tr_{V_A}|\Omega\rangle\otimes\cdots\otimes|\Omega\rangle,
\end{equation}
where $|\Omega\rangle$ denotes the 1-site block, which can be expanded as
\begin{equation}\label{eq:1site f3}
|\Omega\rangle=\omega_i|i\rangle,\quad \omega_i\in {\rm{End}}(V_A).
\end{equation}
Here $V_A$ is an internal space. The integrable state therefore takes the standard matrix-product-state form
\begin{equation}
|\Psi\rangle=\tr_{V_A}(\omega_{i_1}\omega_{i_2}\cdots \omega_{i_N})|i_1,i_2,\cdots,i_N\rangle.
\end{equation}
\subsection{Chiral 1-site states}
The chiral integrability condition (\ref{sec1:f1}) requires $|\Omega\rangle$ to satisfy the constraints
\begin{eqnarray}\label{eq:1site f1}
&&\bar{R}_{02}(u) |\Omega\rangle_2 K_0(u)=K_0(u) R_{02}(-u-2) |\Omega\rangle_2,\\ \label{eq:1site f2}
&&R_{01}(u) |\Omega\rangle_1 K_0(u)=K_0(u) \bar{R}_{01}(-u-2) |\Omega\rangle_1.
\end{eqnarray}
These equations may be regarded as two mixed state-operator equations on ${\rm{End}}(V)\otimes V$. Introducing
\begin{equation}\label{eq:1site f4}
|\psi(u)\rangle=K^{\alpha}_{\beta}(u)|\alpha,\beta\rangle=K^{\alpha}_{\beta}(u)|\alpha\rangle\otimes|\beta\rangle,\quad K^{\alpha}_{\beta}(u)\in {\rm{End}}(V_A),
\end{equation}
we can rewrite (\ref{eq:1site f1}) and (\ref{eq:1site f2}) as pure state equations on $V^{\otimes 3}$:
\begin{eqnarray}
&&\bar{R}_{12}(u)|\Omega\rangle_1\otimes |\psi(u)\rangle_{23}=\bar{R}_{13}(u) |\psi(u)\rangle_{23}\otimes |\Omega\rangle_1,\\
&&{R}_{12}(u)|\Omega\rangle_1\otimes |\psi(u)\rangle_{23}={R}_{13}(u) |\psi(u)\rangle_{23}\otimes |\Omega\rangle_1.
\end{eqnarray}
The corresponding component forms are
\begin{eqnarray}
&&\bar{R}^{i,j}_{\alpha,\beta}(u) \omega_{\beta} K^{\alpha}_{m}(u)=K^{i}_{\alpha}(u) R^{\alpha,j}_{m,\beta}(-u-2)\omega_{\beta},\\
&&{R}^{i,j}_{\alpha,\beta}(u) \omega_{\beta} K^{\alpha}_{m}(u)=K^{i}_{\alpha}(u) \bar{R}^{\alpha,j}_{m,\beta}(-u-2)\omega_{\beta}.
\end{eqnarray}
These equations are the ABJM analogues of the square-root relations introduced in \cite{Pozsgay:2018dzs}. Substituting the explicit expressions for $R(u)$ and $\bar{R}(u)$ in (\ref{sec1:f3}), and using the definitions (\ref{eq:1site f3}) and (\ref{eq:1site f4}), these square-root relations reduce to
\begin{eqnarray}\label{eq:1site f5}
(u+2)\left[\omega_a, K^b_c(u)\right]&=&\delta_{ab}\,\omega_d \,K^d_c(u)-\delta_{ac}K^b_d(u) \,\omega_d,\\ \label{eq:1site f6}
 u\left[\omega_a, K^b_c(u)\right]&=&K^b_a(u)\,\omega_c-\omega_b \,K^a_c(u).
\end{eqnarray}
Since $K^a_b(u)$ and $\omega_a$ act on the same internal space $V_A$, they do not commute in general: $\left[K^a_b(u),\omega_c\right]\neq 0$. If there is no internal space, namely $K^a_b(u)\in \mathbb{C},\,\omega_a \in \mathbb{C}$, then equation (\ref{eq:1site f5}) reduces to
\begin{equation}
0=\delta_{ab} \left[\omega^t K(u)\right]_c-\delta_{ac} \left[K(u)\omega\right]^b.
\end{equation}
Here $\omega=(\omega_1,\omega_2,\omega_3,\omega_4)^t$. Since this equation holds for arbitrary $a,b,c$, we obtain $K(u)\omega=0$. If $K(u)$ is invertible, this further implies $\omega=0$. Therefore, a nontrivial chiral integrable state requires operator-valued solutions $\omega_a\in {\rm{End}}(V_A)$ and $K(u)\in {\rm{End}}(V\otimes V_A)$.

We now focus on equation (\ref{eq:1site f6}) and consider a Clifford-algebra realization of this relation. Such a construction was first introduced in \cite{Pozsgay:2018dzs} for the symmetric pair $(SU(N),SO(N))$. We assume that $\omega_a$ are generators of a four-dimensional Euclidean Clifford algebra:
\begin{equation}
  \{\omega_a,\omega_b\}=2\delta_{ab},\quad a,b=1,\cdots,4.
  \end{equation}
We then make the following ansatz for $K(u)$:
\begin{equation}
  K^a_b(u)=f(u)\delta_{ab}+\omega_a\omega_b.
  \end{equation}
To fix $f(u)$, we substitute this ansatz for $K(u)$ into equation (\ref{eq:1site f6}), which gives $f(u)=2u$. It remains to verify that $K^a_b(u)=2u\delta_{ab}+\omega_a\omega_b$ also satisfies equation (\ref{eq:1site f5}). This follows from the identities
\begin{eqnarray}
  \left[\omega_a,K^b_c(u)\right]=\left[\omega_a,\omega_b\omega_c\right]=\{\omega_a,\omega_b\}\omega_c-\omega_b\{\omega_a,\omega_c\}=2(\delta_{ab}\omega_c-\delta_{ac}\omega_b),
  \end{eqnarray}
\begin{equation}
  \begin{split}
  K^b_a(u)\omega_c-\omega_b K^a_c(u)=&(2u\delta_{ba}+\omega_b\omega_a)\omega_c-\omega_b(2u\delta_{ac}+\omega_a\omega_c)=2u(\delta_{ab}\omega_c-\delta_{ac}\omega_b).\\
  \end{split}
  \end{equation}
Thus we obtain a chiral 1-site integrable state for the ABJM spin chain in terms of Clifford algebra generators. A concrete representation may be chosen as
\begin{equation}
\begin{split}
\omega_1=\sigma_x\otimes \mathbb{I}_2,\quad \omega_2=\sigma_y\otimes\mathbb{I}_2,\quad \omega_3=\sigma_z\otimes\sigma_x,\quad \omega_4=\sigma_z\otimes \sigma_y.
\end{split}
\end{equation}
Finally, one can show that $K(u)$ is invertible. Introducing the operators $\Pi_{ab}$,
\begin{equation}
\Pi_{ab}=\frac{1}{4}\omega_a\omega_b,
\end{equation}
it can be easily shown that $\Pi:=e_{ab}\otimes \Pi_{ab}\in {\rm{End}}(V\otimes V_A)$ is a projector: $\Pi^2=\Pi$ and
the matrix $K(u)$ can be written as
\begin{equation}
 K(u)=2u\mathbb{I}+4 \Pi,
\end{equation}
so its inverse is
\begin{equation}
  K^{-1}(u)=\frac{1}{2u}\mathbb{I}-\frac{1}{u(u+2)}\Pi.
  \end{equation}
\subsection{Achiral 1-site states}
For an achiral 1-site block $|\Omega\rangle$, the achiral $KT$-relation (\ref{sec1:f2}) leads to the following state-operator mixed equations 
\begin{equation}
\begin{split}
\bar{R}_{02}(u) |\Omega\rangle_2 K_0(u)=K_0(u) \bar{R}_{02}(-u-2) |\Omega\rangle_2,\\
R_{01}(u) |\Omega\rangle_1 K_0(u)=K_0(u) R_{01}(-u-2) |\Omega\rangle_1.
\end{split}
\end{equation}
These equations can be converted into the pure state equations on $V^{\otimes 3}$:
\begin{eqnarray}
&&\bar{R}_{12}(u)|\Omega\rangle_1\otimes |\psi(u)\rangle_{23}=R_{13}(u) |\psi(u)\rangle_{23}\otimes |\Omega\rangle_1,\\
&&R_{12}(u)|\Omega\rangle_1\otimes |\psi(u)\rangle_{23}=\bar{R}_{13}(u) |\psi(u)\rangle_{23}\otimes |\Omega\rangle_1.
\end{eqnarray}
Their component forms are
\begin{eqnarray}\label{eq:1site f7}
-(u+2)\omega_a K^b_c(u)+\delta_{ab}\omega_d K^d_c(u)=u K^b_c(u) \omega_a+K^b_a(u) \omega_c,\\ \label{eq:1site f8}
u \omega_a K^b_c(u) +\omega_b K^a_c(u)=-(u+2)K^b_c(u) \omega_a+\delta_{ac} K^b_d(u) \omega_d.
\end{eqnarray}
We introduce the following six contractions between $\omega_a$ and $K^c_d(u)$:
\begin{equation}
\begin{split}
&L_a(u):=\omega_d K^d_a(u),\quad U^a(u):=\omega_d K^a_d(u),\quad X_a(u):=\omega_a K^d_d(u),\\
&S_a(u):=K^d_a(u) \omega_d,\quad R^a(u):=K^a_d(u) \omega_d,\quad \,Y_a(u):=K^d_d(u) \omega_a.
\end{split}
\end{equation}
Contracting the three pairs of indices $(a,b)$, $(a,c)$ and $(b,c)$ separately in equation (\ref{eq:1site f7}) gives
\begin{eqnarray}
&&(u-2)L_a(u)+uS_a(u)+Y_a(u)=0,\\
&&L_a(u)-(u+2)U^a(u)-(u+1)R^a(u)=0,\\
&&L_a(u)-(u+2)X_a(u)-S_a(u)-u Y_a(u)=0.
\end{eqnarray}
Similarly, contracting $(a,b)$, $(a,c)$ and $(b,c)$ separately in equation (\ref{eq:1site f8}) gives
\begin{eqnarray}
&&(u+1)L_a(u)+(u+2)S_a(u)-R^a(u)=0,\\
&&uU^a(u)+X_a(u)+(u-2)R^a(u)=0,\\
&&U^a(u)+u X_a(u)-R^a(u)+(u+2)Y_a(u)=0.
\end{eqnarray}
These six equations can be written in matrix form as
\begin{equation}\label{eq:1site f9}
\begin{pmatrix}
2-u & -u & 0 & 0 & 0 & -1\\
1 & 0 & -(u+1) & -(u+2) & 0 & 0\\
1 & -1 & 0 & 0 & -(u+2) & -u\\
0 & 0 & 2-u & -u & -1 & 0\\
-(u+1) & -(u+2) & 1 & 0 & 0 & 0\\
0 & 0 & 1 & -1 & -u & -(u+2)
\end{pmatrix}
\begin{pmatrix}
L_a(u)\\
S_a(u)\\
R^a(u)\\
U^a(u)\\
X_a(u)\\
Y_a(u)
\end{pmatrix}
=0.
\end{equation}
The determinant of the coefficient matrix is $-36(u+1)(u+2)$, which is nonzero for generic $u$. Hence equation (\ref{eq:1site f9}) has only the trivial solution. In particular,
\begin{equation}
R^a(u)=(K(u)\omega)^a=0\quad \Rightarrow \quad K(u)\,\omega=0,
\end{equation}
and, if $K(u)$ is invertible, $\omega=0$.

\section{Integrable 2-site states}
In this section, we study integrable boundary states with 2-site translational invariance. Such states are constructed from a basic 2-site block as
\begin{equation}
|\Psi\rangle=|\phi\rangle_{12}\otimes |\phi\rangle_{34}\otimes\cdots\otimes |\phi\rangle_{2N-1,2N},
\end{equation}
where we assume that the 2-site block contains no internal hidden space: $|\phi\rangle\in V\otimes V$. The 2-site block can be expanded as
\begin{equation}
|\phi\rangle=M^i_j|i\rangle\otimes |{j}\rangle,\quad M^i_j\in \mathbb{C}.
\end{equation}
\subsection{Chiral 2-site states}
We first consider the chiral 2-site state. The chiral $KT$-relation imposes the following condition on the 2-site block:
\begin{equation}\label{eq:2site f1}
R_{01}(u)\bar R_{02}(u)|\phi\rangle_{12}K_0(u)=K_0(u)\bar{R}_{01}(-u-2)R_{02}(-u-2)|\phi\rangle_{12},
\end{equation}
which is a mixed state-operator equation on ${\rm{End}}(V)\otimes V\otimes V$. We now transform it into a pure operator equation on ${\rm{End}}(V\otimes V)$. To this end, we first let both sides of equation (\ref{eq:2site f1}) act on a basis vector of $V_0$, obtaining
\begin{equation}
\begin{split}
\Big[R_{01}(u)\bar R_{02}(u)|\phi\rangle_{12}K_0(u)\Big]|\alpha\rangle_0
&=\Big[R_{01}(u)\bar R_{02}(u)M^i_jK_0(u)\Big]|\alpha,i,{j}\rangle_{012} \\
&=\Big[R^{\beta k}_{\alpha_1 i}(u)M^i_j
R^{\alpha_1{j}}_{\alpha_2\ell}(-u-2)K^{\alpha_2}_{\alpha}(u)\Big]
|\beta,k,{\ell}\rangle_{012},
\end{split}
\end{equation}
and
\begin{equation}
\begin{split}
&\Big[K_0(u)R_{01}(-u-2)\bar R_{02}(-u-2)|\phi\rangle_{12}\Big]|\alpha\rangle_0 \\
=&\Big[K_0(u)R_{01}(-u-2)\bar R_{02}(-u-2)M^i_j\Big]|\alpha,i,{j}\rangle_{012} \\
=&\Big[K^{\beta}_{\alpha_1}(u)\bar{R}^{\alpha_1 k}_{\alpha_2 i}(-u-2)M^i_j
\bar{R}^{\alpha_2{j}}_{\alpha\ell}(u)\Big]
|\beta,k,{\ell}\rangle_{012},
\end{split}
\end{equation}
where we have used the basic relation
\begin{equation}
\bar{R}^{ij}_{kl}(u)=R^{il}_{kj}(-u-2).
\end{equation}
Thus we obtain the component equality
\begin{equation}
\begin{split}
&R^{\beta k}_{\alpha_1 i}(u)M^i{}_j
R^{\alpha_1{j}}_{\alpha_2\ell}(-u-2)K^{\alpha_2}_{\alpha}(u) \\
&\qquad =K^{\beta}_{\alpha_1}(u)\bar{R}^{\alpha_1 k}_{\alpha_2 i}(-u-2)M^i_j
\bar{R}^{\alpha_2{j}}_{\alpha\ell}(u),
\end{split}
\end{equation}
which may be regarded as an operator equation on $V\otimes V$:
\begin{equation}\label{eq:2site f2}
R_{12}(u)M_1R_{12}(-u-2)K_2(u)=K_2(u)\bar{R}_{12}(-u-2)M_1\bar{R}_{12}(u).
\end{equation}
We now show that the above equation admits only the trivial solution for the integrable pair $(M,K(u))$. The proof is simple. At the special point $u=-2$, equation (\ref{eq:2site f2}) reduces to
\begin{equation}\label{eq:2site f3}
R_{12}(-2)M_1R_{12}(0)K_2(-2)=K_2(-2)\bar{R}_{12}(0)M_1 \mathbb{Q}_{12}.
\end{equation}
Since $R_{12}(-2)$, $R_{12}(0)$ and $\bar{R}_{12}(0)$ are invertible matrices on $V\otimes V$, and since we require $K(u)$ to be invertible for generic $u$, equation (\ref{eq:2site f3}) implies the relation
\begin{equation}
{\rm{\mathbf{rank}}}(M\otimes \mathbb{I})={\rm{\mathbf{rank}}}(M_1 \mathbb{Q}_{12}).
\end{equation}
However, the ranks of the two sides can be evaluated as
\begin{eqnarray}
{\rm{\mathbf{rank}}}(M\otimes \mathbb{I})=4{\rm{\mathbf{rank}}}(M),\\
{\rm{\mathbf{rank}}}(M_1 \mathbb{Q}_{12})\leq {\rm{\mathbf{rank}}}(\mathbb{Q}_{12})=1,
\end{eqnarray}
and it follows that ${\rm{\mathbf{rank}}}(M)=0$. Hence only the trivial solution $M=0$ exists.

\subsection{Achiral 2-site states}
In the achiral case, the 2-site state satisfies
\begin{equation}
R_{01}(u)\bar R_{02}(u)|\phi\rangle_{12}K_0(u)=K_0(u)R_{01}(-u-2)\bar R_{02}(-u-2)|\phi\rangle_{12},
\end{equation}
which can be converted into the following operator equation on $V\otimes V$:
\begin{equation}\label{eq:2site f4}
R_{12}(u)M_1R_{12}(-u-2)K_2(u)=K_2(u)R_{12}(-u-2)M_1R_{12}(u).
\end{equation}
We now solve for the achiral integrable pair $(M,K(u))$ as a solution of the above equation. We do not assume in advance that $K(u)$ is a solution of the SP-type reflection equation.

Substituting the explicit expression for the $R$-matrix into (\ref{eq:2site f4}), we obtain
\begin{equation}\label{eq:2site f5}
\begin{split}
u\Big[M_1 K_1(u)-K_2(u) M_2\Big]\mathbb{P}_{12}+(u+2)\Big[K_2(u) M_1-M_2 K_1(u)\Big]\mathbb{P}_{12}\\+M_2 K_2(u)-K_2(u) M_2=0.
\end{split}
\end{equation}
Taking the partial trace over $V_2$ in equation (\ref{eq:2site f5}) and using the simple identities
\begin{eqnarray}
&&\tr_2(M_1 K_1(u)\mathbb{P}_{12})=\tr_2(K_2(u) M_1\mathbb{P}_{12})=M K(u),\\
&&\tr_2(K_2(u) M_2\mathbb{P}_{12})=\tr_2(M_2 K_1(u)\mathbb{P}_{12})=K(u) M,
\end{eqnarray}
we find
\begin{equation}
\left[M,K(u)\right]=0.
\end{equation}
Equation (\ref{eq:2site f5}) then becomes
\begin{equation}\label{eq:2site f6}
u\Big[M_1 K_1(u)-K_2(u) M_2\Big]+(u+2)\Big[K_2(u) M_1-M_2 K_1(u)\Big]=0.
\end{equation}
Taking the partial trace over $V_1$ in this equation gives
\begin{equation}
\begin{split}
MK(u)=&\frac{\tr(MK(u))}{n}\,\mathbb{I}-\left(\frac{u+2}{nu}\tr K(u)\right) M+\left(\frac{u+2}{nu}\tr M\right)K(u)\\
:=&a(u) \mathbb{I}+b(u) M+c(u) K(u).
\end{split}
\end{equation}
Substituting this relation back into equation (\ref{eq:2site f6}), we obtain
\begin{equation}\label{eq:2site f7}
\begin{split}
ub(u)\left(M\otimes \mathbb{I}-\mathbb{I}\otimes M\right)+uc(u)\left(K(u)\otimes \mathbb{I}-\mathbb{I}\otimes K(u)\right)
\\+(u+2)\left(M\otimes K(u)-K(u)\otimes M\right)=0.
\end{split}
\end{equation}
It remains to distinguish the following two cases:

\noindent $\bullet$  $M=c\mathbb{I}$: In this case, equation (\ref{eq:2site f6}) gives $K_1(u)=K_2(u)$, which shows that $K(u)$ is a scalar matrix, $K(u)=k(u)\mathbb{I}$. The corresponding local 2-site block is
      \begin{equation}
  |\phi\rangle=M^i_j|i,j\rangle\sim |1,1\rangle+|2,2\rangle+\cdots +|n,n\rangle,
  \end{equation}
  which is usually called the ``$\delta$'' state in the literature.

\noindent $\bullet$  $M\neq c\mathbb{I}$: In this case, equation (\ref{eq:2site f7}) implies
  \begin{equation}
  K(u)=\alpha(u)\mathbb{I}+\beta(u)M.
  \end{equation}
  Substituting this ansatz for $K(u)$ back into the full equation (\ref{eq:2site f6}), we obtain
  \begin{equation}
  u\beta(u)M_1^2+(2u+2)\alpha(u)M_1=u\beta(u)M_2^2+(2u+2)\alpha(u)M_2,
  \end{equation}
  which shows that
  \begin{equation}
  u\beta(u)M^2+(2u+2)\alpha(u)M=\gamma(u)\mathbb{I}.
  \end{equation}
  Since $M$ is independent of $u$, we obtain
  \begin{equation}
  \frac{(2u+2)\alpha(u)}{u\beta(u)}=c_1,\quad \frac{\gamma(u)}{u\beta(u)}=c_2,
  \end{equation}
  where $c_1$ and $c_2$ are constants. Therefore, for any second-order constant coefficient matrix equation satisfied by $M$, for example,
  \begin{equation}\label{eq:2site f8}
  M^2+a M+b\mathbb{I}=0,
  \end{equation}
  the ratio $\alpha(u)/\beta(u)$ must satisfy
  \begin{equation}
  \frac{\alpha(u)}{\beta(u)}=a\frac{u}{2u+2},
  \end{equation}
  and the corresponding $K(u)$, up to an overall scalar factor, is given by
  \begin{equation}\label{eq:2site f9}
  K(u)=a u \mathbb{I}+(2u+2) M.
  \end{equation}
  The matrices $M$ and $K(u)$ therefore form an integrable pair satisfying the achiral operator equation (\ref{eq:2site f4}). However, if we define
  \begin{equation}
  G:=M+\frac{a}{2}\mathbb{I},
  \end{equation}
  then equation (\ref{eq:2site f8}) gives
  \begin{equation}
G^2=\Delta \mathbb{I},\quad \Delta=\frac{a^2}{4}-b,
\end{equation}
and $K(u)$ in equation (\ref{eq:2site f9}) becomes
\begin{equation}
K(u)=au \mathbb{I}+(2u+2)(G-\frac{a}{2}\mathbb{I})=-a\mathbb{I}+2(u+1)G.
\end{equation}
Thus we find $K(u)$ is precisely of the SP-type solution described in appendix~\ref{app:SP}.

\section{Integrable 3-site states}
In this section we study integrable 3-site states. We take the total chain length to be $6N$, so that the state contains $2N$ elementary 3-site blocks:
\begin{equation}
|\Psi\rangle=\tr_{V_A}|\Theta\rangle_{123}\otimes |\Theta\rangle_{456}\otimes\cdots \otimes |\Theta\rangle_{6N-2,6N-1,6N},
\end{equation}
where the basic 3-site block $|\Theta\rangle$ is defined as
\begin{equation}
|\Theta\rangle=\omega_{b_1b_2b_3}|b_1,b_2,b_3\rangle:=\omega_{\vec{B}}|\vec{B}\rangle,\quad \omega_{\vec{B}}\in {\rm{End}}(V_A).
\end{equation}
Thus,
\begin{equation}
|\Psi\rangle=\tr_{V_A}(\omega_{\vec{B}_1}\omega_{\vec{B}_2}\cdots \omega_{\vec{B}_{2N}})|\vec{B}_1,\vec{B}_2,\cdots,\vec{B}_{2N}\rangle.
\end{equation}
\subsection{Chiral 3-site states}
To construct chiral integrable 3-site states, the basic block $|\Theta\rangle$ must satisfy the following constraints:
\begin{equation}\label{eq:3site f1}
\begin{split}
\bar{R}_{04}(u)R_{05}(u)\bar{R}_{06}(u)|\Theta\rangle_{456} K_0(u)=K_0(u)R_{04}(-u-2)\bar{R}_{05}(-u-2)R_{06}(-u-2)|\Theta\rangle_{456},\\
{R}_{01}(u)\bar{R}_{02}(u){R}_{03}(u)|\Theta\rangle_{123} K_0(u)=K_0(u)\bar{R}_{01}(-u-2){R}_{02}(-u-2)\bar{R}_{03}(-u-2)|\Theta\rangle_{123}.
\end{split}
\end{equation}
We may define
\begin{eqnarray}
R_{0,123}(u):=R_{01}(u)\bar{R}_{02}(u)R_{03}(u),\quad
\bar{R}_{0,123}(u):=\bar{R}_{01}(u)R_{02}(u)\bar{R}_{03}(u),
\end{eqnarray}
then the component forms of (\ref{eq:3site f1}) are
\begin{eqnarray}
&&\bar{R}^{i,\vec{J}}_{\alpha,\vec{B}}(u)\omega_{\vec{B}}K^{\alpha}_m(u)=K^i_{\alpha}(u)R^{\alpha,\vec{J}}_{m,\vec{B}}(-u-2)\omega_{\vec{B}},\\
&&{R}^{i,\vec{J}}_{\alpha,\vec{B}}(u)\omega_{\vec{B}}K^{\alpha}_m(u)=K^i_{\alpha}(u)\bar{R}^{\alpha,\vec{J}}_{m,\vec{B}}(-u-2)\omega_{\vec{B}},
\end{eqnarray}
from which we obtain the state equations on $V^{\otimes 5}$:
\begin{eqnarray}\label{eq:3site f2}
&&\bar{R}_{14}(u)R_{24}(u)\bar{R}_{34}(u)|\Theta\rangle_{123}\otimes |\psi(u)\rangle_{45}=\bar{R}_{35}(u)R_{25}(u)\bar{R}_{15}(u)|\psi(u)\rangle_{45}\otimes|\Theta\rangle_{123},\\
&&{R}_{14}(u)\bar{R}_{24}(u){R}_{34}(u)|\Theta\rangle_{123}\otimes |\psi(u)\rangle_{45}={R}_{35}(u)\bar{R}_{25}(u){R}_{15}(u)|\psi(u)\rangle_{45}\otimes|\Theta\rangle_{123},
\end{eqnarray}
where, as in the 1-site case, we define $|\psi(u)\rangle:=K^i_j(u)|i,j\rangle$.

Solving for the integrable pair $(|\Theta\rangle,\,K(u))$ from the above two equations directly is technically involved.  We therefore restrict to $c$-number solutions without an internal space $V_A$: $\omega_{abc},K^i_j(u) \in \mathbb{C}$ and prove that, when $K(u)$ satisfies the SNP-type reflection equation, $|\Theta\rangle$ has only the trivial zero solution.

We start from equation (\ref{eq:3site f2}) and consider its reduced form at the degenerate point $u=-2$ of $\bar{R}(u)$:
\begin{equation}
\mathbb{Q}_{14} R_{24}(-2) \mathbb{Q}_{34}|\Theta\rangle_{123}\otimes |\psi(-2)\rangle_{45}=\mathbb{Q}_{35} R_{25}(-2) \mathbb{Q}_{15}|\Theta\rangle_{123}\otimes |\psi(-2)\rangle_{45}.
\end{equation}
By the SNP classification of appendix~\ref{app:SNP}, the relevant solution is a constant symmetric or antisymmetric matrix.  We therefore write $|\psi(u)\rangle=K^i_j|i\rangle\otimes|j\rangle$ with $K$ a constant $c$-number matrix, and then the component form of the above equation is
\begin{equation}\label{eq:3site f3}
\begin{split}
-2\delta_{ps} \omega_{rqc} K^c_t+\delta_{ps}\delta_{qr}\omega_{aac}K^c_t
=-2\delta_{rt}\omega_{aqp}K^s_a+\delta_{rt}\delta_{pq}\omega_{abb}K^s_a.
\end{split}
\end{equation}
We define four matrices $\Omega_b$, $b=1,2,3,4$, by
\begin{equation}
(\Omega_b)^a_c:=\omega_{abc},
\end{equation}
and rewrite equation (\ref{eq:3site f3}) as
\begin{equation}\label{eq:3site f4}
-2\delta_{ps}(\Omega_q K)^r_t+\delta_{ps}\delta_{qr}(\Omega_a K)^a_t=-2\delta_{rt}(K \Omega_q)^s_p+\delta_{rt}\delta_{pq}(K\Omega_b)^s_b.
\end{equation}
Letting $p=s,\,r\neq t$ in equation (\ref{eq:3site f4}), we obtain
\begin{equation}\label{eq:3site nd1}
(\Omega_a K)^b_c=\frac{1}{2}\delta_{ab}(\Omega_d K)^d_c,\quad b\neq c.
\end{equation}
Letting $p\neq s,\,r=t$ in equation (\ref{eq:3site f4}), we obtain
\begin{equation}\label{eq:3site nd2}
(K \Omega_a)^b_c=\frac{1}{2}\delta_{ac}(K \Omega_d)^b_d,\quad b\neq c.
\end{equation}
The above two equations give the non-diagonal elements of the matrices $\Omega_a K$ and $K\Omega_a$, respectively. To find the diagonal elements, we let $p=s,\,r=t$ in equation (\ref{eq:3site f4}) to obtain
\begin{equation}\label{eq:3site f5}
-2(\Omega_a K)^i_i+\delta_{ai}(\Omega_d K)^d_i=-2(K\Omega_a)^j_j+\delta_{aj}(K \Omega_d)^j_d,
\end{equation}
where the $i,j$ indices are not summed. Since equation (\ref{eq:3site f5}) holds for arbitrary $i$ and $j$, both sides of the equation must be equal to the same value, which depends only on the index $a$,
\begin{eqnarray}
-2(\Omega_a K)^i_i+\delta_{ai}(\Omega_d K)^d_i=\lambda_a,\quad
-2(K\Omega_a)^j_j+\delta_{aj}(K \Omega_d)^j_d=\lambda_a,
\end{eqnarray}
from which we obtain the diagonal elements:
\begin{eqnarray}\label{eq:3site diag}
(\Omega_a K)^b_b=\frac{1}{2}\Big[\delta_{ab}(\Omega_d K)^d_b-\lambda_a\Big],\quad
(K \Omega_a)^b_b=\frac{1}{2}\Big[\delta_{ab}(K \Omega_d)^b_d-\lambda_a\Big],
\end{eqnarray}
where the index $b$ is not summed. We introduce two column vectors $[x]$ and $[y]$ defined as
\begin{equation}\label{eq:3site de}
x_a:=(\Omega_d K)^d_a,\quad y_a:=(K \Omega_d)^a_d.
\end{equation}
Combining the off-diagonal results (\ref{eq:3site nd1},\,\ref{eq:3site nd2}) and the diagonal expression (\ref{eq:3site diag}), we find
\begin{eqnarray}\label{eq:3site ko}
\Omega_a K=-\frac{1}{2}\lambda_a \mathbb{I}+\frac{1}{2}e_a x^t,\quad
K \Omega_a=-\frac{1}{2}\lambda_a \mathbb{I}+\frac{1}{2}y e_a^t,
\end{eqnarray}
where $e_a$ is the standard basis vector and the symbol `t' denotes the transposition. For an invertible $K$-matrix, we have
\begin{equation}
K\Omega_a=K(\Omega_a K)K^{-1}\quad \Rightarrow \quad y e_a^t=(K e_a) (x^t K^{-1}).
\end{equation}
Since $y e_a^t$ has nonzero entries only in the $a$-th column, the row vector $x^t K^{-1}$ can have a nonzero component only in the $a$-th position.  Since the index `$a$' is arbitrary, the only possibility is $x=0$, and consequently $y=0$. In this case, equation (\ref{eq:3site ko}) becomes
\begin{equation}
\Omega_a K=-\frac{1}{2}\lambda_a \mathbb{I}
\end{equation}
Then the defining equation (\ref{eq:3site de}) gives
\begin{equation}
x_a=-\frac{1}{2}\lambda_d\delta_{da}=-\frac{1}{2}\lambda_a=0,
\end{equation}
which implies $\Omega_a K=0$. We therefore conclude that $\Omega_a=0$, or $\omega_{abc}=0$, showing that the chiral integrable counterpart of a constant $K$-matrix can only be the zero 3-site state $|\Theta\rangle=0$.

\subsection{Achiral 3-site states}
In the achiral case, the basic 3-site block must satisfy
\begin{align}
&\bar{R}_{04}(u)R_{05}(u)\bar{R}_{06}(u)|\Theta\rangle_{456} K_0(u) \notag\\
&\qquad =
K_0(u)\bar{R}_{04}(-u-2)R_{05}(-u-2) \bar{R}_{06}(-u-2)|\Theta\rangle_{456},
\\
&{R}_{01}(u)\bar{R}_{02}(u){R}_{03}(u)|\Theta\rangle_{123} K_0(u) \notag\\
&\qquad =
K_0(u){R}_{01}(-u-2)\bar{R}_{02}(-u-2) {R}_{03}(-u-2)|\Theta\rangle_{123}.
\end{align}
The component forms are
\begin{eqnarray}
&&\bar{R}^{i,\vec{J}}_{\alpha,\vec{B}}(u)\omega_{\vec{B}}K^{\alpha}_m(u)=K^i_{\alpha}(u)\bar{R}^{\alpha,\vec{J}}_{m,\vec{B}}(-u-2)\omega_{\vec{B}},\\
&&{R}^{i,\vec{J}}_{\alpha,\vec{B}}(u)\omega_{\vec{B}}K^{\alpha}_m(u)=K^i_{\alpha}(u)R^{\alpha,\vec{J}}_{m,\vec{B}}(-u-2)\omega_{\vec{B}},
\end{eqnarray}
from which we obtain the equivalent state equation on $V^{\otimes 5}$:
\begin{eqnarray}
\bar{R}_{14}(u)R_{24}(u)\bar{R}_{34}(u)|\Theta\rangle_{123}\otimes |\psi(u)\rangle_{45}=
{R}_{35}(u)\bar{R}_{25}(u){R}_{15}(u)|\psi(u)\rangle_{45}\otimes|\Theta\rangle_{123},\\ \label{eq:3site f6}
{R}_{14}(u)\bar{R}_{24}(u){R}_{34}(u)|\Theta\rangle_{123}\otimes |\psi(u)\rangle_{45}=\bar{R}_{35}(u)R_{25}(u)\bar{R}_{15}(u)|\psi(u)\rangle_{45}\otimes|\Theta\rangle_{123}.
\end{eqnarray}
We now specialize $K(u)$ to the SP-type solution $K(u)=\xi \mathbb{I}+u G$ classified in appendix~\ref{app:SP}, where $G^2=0$ or $G^2=\mathbb{I}$. We also require $K(u)$ to be invertible. Therefore, if $\xi\neq 0$, $K(0)=\xi \mathbb{I}$; if $\xi=0$, we can only choose $G^2=\mathbb{I}$, and then $K(u)=u G$. By discarding the overall factor $u$, we have $K(0)=G$. Thus, in either case, we may restrict $K(u)$ at the point $u=0$ to the form
\begin{equation}
[K(0)]^a_b:=K^a_b=\kappa_a\delta_{ab},\quad \kappa_a\neq 0.
\end{equation}
The equation (\ref{eq:3site f6}) at $u=0$ reduces to
\begin{equation}
\mathbb{P}_{14}(-2+\mathbb{Q}_{24})\mathbb{P}_{34}|\Theta\rangle_{123}\otimes |\psi(0)\rangle_{45}=(-2+\mathbb{Q}_{35})\mathbb{P}_{25}(-2+\mathbb{Q}_{15})|\Theta\rangle_{123}\otimes |\psi(0)\rangle_{45},
\end{equation}
whose component form is
\begin{equation}
\begin{split}
\left(-2\omega_{sqp}+\delta_{pq}\omega_{sbb}\right)K^r_t
=4\omega_{ptr}K^s_q
-2\delta_{rt}\omega_{pbb}K^s_q
-2\delta_{pq}\omega_{atr}K^s_a
+\delta_{pq}\delta_{rt}\omega_{abb}K^s_a.
\end{split}
\end{equation}
Substituting $K^{a}_b=\kappa_a\delta_{ab}$ into the above equation, we obtain
\begin{equation}\label{eq:3site c1}
\begin{split}
\kappa_r\delta_{rt}\left(-2\omega_{sqp}+\delta_{pq}\omega_{sbb}\right)
=4\kappa_s\delta_{sq}\omega_{ptr}
-2\delta_{rt}\kappa_s\delta_{sq}\omega_{pbb}
-2\delta_{pq}\kappa_s\omega_{str}
+\delta_{pq}\delta_{rt}\kappa_s\omega_{sbb}.
\end{split}
\end{equation}
Setting $p\neq q,\,r\neq t,\,q=s$ in equation (\ref{eq:3site c1}), we find $\kappa_q \omega_{ptr}=0$, which implies
\begin{equation}\label{eq:3site c2}
\omega_{abc}=0,\quad b\neq c.
\end{equation}
Setting $p\neq q,\,r= t,\,q=s$ in (\ref{eq:3site c1}), we obtain
\begin{equation}\label{eq:3site c3}
-2\kappa_r\omega_{qqp}=4\kappa_q \omega_{prr}-2\kappa_q \omega_{pbb},
\end{equation}
where the indices $p$ and $r$ are not summed. Combining equations (\ref{eq:3site c2}) and (\ref{eq:3site c3}), we find
\begin{equation}
2\omega_{prr}=\sum_b\omega_{pbb}\,\Rightarrow\,2\sum_r\omega_{prr}=4\sum_b\omega_{pbb}\,\Rightarrow\,\sum_b\omega_{pbb}=0\,\Rightarrow\,\omega_{prr}=0.
\end{equation}
Therefore $\omega_{abc}\equiv 0,\,\forall a,b,c$, and hence $|\Theta\rangle=0$.

\section{Integrable 4-site states}
In this section we investigate integrable 4-site states.  As in the 2-site case, we convert the basic mixed state-operator equations into purely algebraic operator equations and then solve them using group-theoretic methods under certain assumptions.  We focus on $c$-number integrable product states without internal spaces, expressed in terms of 4-site blocks as follows:
\begin{equation}
 |\Psi\rangle=|\phi\rangle_{1234}\otimes\cdots\otimes |\phi\rangle_{4N-3,4N-2,4N-1,4N}.
\end{equation}
\subsection{Chiral 4-site states}
The basic state-operator equation on ${\rm{End}}(V)\otimes V^{\otimes 4}$ for a chiral 4-site block $|\phi\rangle$ is
\begin{equation}\label{eq:4site state-oper}
\begin{split}
&R_{01}(u)\bar{R}_{02}(u)R_{03}(u)\bar{R}_{04}(u)|\phi\rangle_{1234}K_0(u)\\
=&K_0(u)\bar{R}_{01}(-u-2)R_{02}(-u-2)\bar{R}_{03}(-u-2)R_{04}(-u-2)|\phi\rangle_{1234}.
\end{split}
\end{equation}
We define
\begin{equation}
|\phi\rangle_{1234}=M^{i,j}_{k,l}|i,j,k,l\rangle_{1234},\quad M^{i,j}_{k,l}:=\langle i,j|\mathbb{M}_{12}|k,l\rangle,
\end{equation}
and then, by acting with both sides of equation (\ref{eq:4site state-oper}) on an arbitrary basis vector $|\alpha\rangle$ in the space $V_0$, we obtain
\begin{equation}
\begin{split}
&R_{01}(u)\bar{R}_{02}(u)R_{03}(u)\bar{R}_{04}(u)|\phi\rangle_{1234}K_0(u)|\alpha\rangle_0\\
=& R^{\beta,i'}_{a_1,i}(u)\bar{R}^{a_1,j'}_{a_2,j}(u)R^{a_2,k'}_{a_3,k}(u)\bar{R}^{a_3,l'}_{a_4,l}(u)M^{i,j}_{k,l}K^{a_4}_{\alpha}(u)|\beta,i',j',k',l'\rangle\\
=&\Big[R_{31}(u)\bar{R}_{32}(u)\mathbb{M}_{12}\bar{R}_{31}(-u-2)R_{32}(-u-2)K_3(u)\Big]^{i',j',\beta}_{k',l',\alpha}|\beta,i',j',k',l'\rangle,
\end{split}
\end{equation}
and
\begin{equation}
\begin{split}
&K_0(u)\bar{R}_{01}(-u-2)R_{02}(-u-2)\bar{R}_{03}(-u-2)R_{04}(-u-2)|\phi\rangle_{1234}|\alpha\rangle_0\\
=&K^{\beta}_{a_1}(u)\bar{R}^{a_1,i'}_{a_2,i}(-u-2){R}^{a_2,j'}_{a_3,j}(-u-2)\bar{R}^{a_3,k'}_{a_4,k}(-u-2){R}^{a_4,l'}_{\alpha,l}(-u-2)M^{i,j}_{k,l}|\beta,i',j',k',l'\rangle\\
=&\Big[K_3(u)\bar{R}_{31}(-u-2)R_{32}(-u-2)\mathbb{M}_{12}R_{31}(u)\bar{R}_{32}(u)\Big]^{i',j',\beta}_{k',l',\alpha}|\beta,i',j',k',l'\rangle.
\end{split}
\end{equation}
From these results we obtain the operator equation on ${\rm{End}}(V^{\otimes 3})$ for the chiral 4-site integrable pair $(\mathbb{M},K(u))$,
\begin{equation}\label{eq:4site-central}
\begin{split}
&R_{13}(u)\bar{R}_{23}(u)\mathbb{M}_{12}\bar{R}_{13}(-u-2)R_{23}(-u-2)K_3(u)\\
=&K_3(u)\bar{R}_{13}(-u-2)R_{23}(-u-2)\mathbb{M}_{12}R_{13}(u)\bar{R}_{23}(u).
\end{split}
\end{equation}
We investigate the $c$-number solutions $(\mathbb M,K(u))$ of this equation without assuming \textit{a priori} that $K(u)$ satisfies the SNP-type reflection equation. We only require $K(u)$ to be invertible for generic $u$ and to admit the Laurent-type expansion,
\begin{equation}
K(u)=K^{(0)}+\frac{K^{(1)}}{u}+\frac{K^{(2)}}{u^2}+\mathcal{O}(u^{-3})\qquad \det K^{(0)}\neq0.
\end{equation}
Substituting the above $K(u)$ into (\ref{eq:4site-central}), and letting $u\rightarrow \infty$, we find that the $u^{-4}$-order terms cancel, while the next $u^{-3}$ order gives
\begin{equation}\label{eq:4site-u3}
\Big[(\mathbb{Q}_{23}-\mathbb{P}_{13})\mathbb{M}_{12}+\mathbb{M}_{12}(\mathbb{P}_{23}-\mathbb{Q}_{13})\Big]K_3^{(0)}=
K_3^{(0)}\Big[\mathbb{M}_{12}(\mathbb{Q}_{23}-\mathbb{P}_{13})+(\mathbb{P}_{23}-\mathbb{Q}_{13})\mathbb{M}_{12}\Big].
\end{equation}
Introducing
\begin{equation}
G_{ab}:=(K^{(0)})^t E_{ab}-K^{(0)} E_{ba},
\end{equation}
and extracting the component of equation (\ref{eq:4site-u3}) on $(E_{ab})_3$, we obtain
\begin{equation}\label{eq:4site commutant}
\Big[\mathbb{M}_{12}\mathbb{P}_{12}, G_{ab}\otimes \mathbb{I}-\mathbb{I}\otimes G^t_{ab}\Big]=0,\quad  a,b=1,\cdots,4.
\end{equation}
This commutant structure is the key reason why a group-theoretic analysis is applicable: if $\{G_{ab}\}$ form a representation of a Lie algebra $\mathfrak{g}$, then $\{-G_{ab}^t\}$ is the corresponding contragredient action, and equation (\ref{eq:4site commutant}) indicates that $\mathbb{M}_{12}\mathbb{P}_{12}$ commutes with the diagonal action of $\mathfrak{g}$. The type of algebra $\mathfrak{g}$ is determined by the concrete choice of $K^{(0)}$. We now study three concrete Lie-algebra realizations of $G_{ab}$.

\subsubsection{$\mathfrak{so}(4)$ case: $(K^{(0)})^t=K^{(0)}$}
For any symmetric invertible $K^{(0)}$, there exists an invertible matrix $A$ such that
\begin{equation}
 A^tK^{(0)}A=\mathbb {I}.
\end{equation}
Moreover, in this case
\begin{equation}
  \begin{split}
  G_{ab}\otimes \mathbb{I}-\mathbb{I}\otimes G^t_{ab}=K^{(0)}(E_{ab}-E_{ba})\otimes \mathbb{I}+\mathbb{I}\otimes (E_{ab}-E_{ba})K^{(0)}.
  \end{split}
  \end{equation}
Thus, according to the formula given in appendix~\ref{appx:matrix}, we have
\begin{equation}\label{eq:4site so4}
  S^{-1}_{12}(G_{ab}\otimes \mathbb{I}-\mathbb{I}\otimes G^t_{ab})S_{12}=F_{ab}\otimes \mathbb{I}+\mathbb{I}\otimes F_{ab},
\end{equation}
where
\begin{eqnarray}
&&S_{12}=A^{-t}\otimes A,\\
&&F_{ab}=A^{-1}(E_{ab}-E_{ba})A^{-t},
\end{eqnarray}
and we denote $A^{-t}=(A^t)^{-1}$. It is clear that the algebra generated by $\{F_{ab}\}$ is isomorphic to $\mathfrak{so}(4)$. We also have
\begin{equation}\label{eq:4site Mhat}
  \begin{split}
  S^{-1}_{12}(\mathbb{M}_{12}\mathbb{P}_{12})S_{12}=(A^t\otimes A^{-1}) \mathbb{M}_{12}(A\otimes A^{-t}) \mathbb{P}_{12},
  \end{split}
  \end{equation}
Thus, if we write
\begin{equation}
\widehat{\mathbb{M}}_{12}=(S_L)_{12}\mathbb{M}_{12} (S_R)_{12},\qquad (S_L)_{12}=A^t\otimes A^{-1},\quad(S_R)_{12}=A\otimes A^{-t},
\end{equation}
then from equations (\ref{eq:4site so4}) and (\ref{eq:4site Mhat}), we find
\begin{equation}
\Big[\widehat{\mathbb{M}}_{12}\mathbb{P}_{12}, X\otimes \mathbb{I}+\mathbb{I}\otimes X\Big]=0,\quad \forall X\in \mathfrak{so}(4),
\end{equation}
showing that
\begin{equation}
\widehat{\mathbb{M}}_{12}\mathbb{P}_{12}\in {\rm{End}}_{\mathfrak{so}(4)}(V\otimes V).
\end{equation}
The irreducible decomposition of $\mathfrak{so}(4)$ diagonal action on $V\otimes V$ is
\begin{equation}
  V\otimes V=W_0\oplus W_s\oplus W_+\oplus W_-\quad \Leftrightarrow\quad \mathbf{4}\otimes \mathbf{4}=\mathbf{1}\oplus \mathbf{9}\oplus \mathbf{3}\oplus \mathbf{3},
  \end{equation}
where $W_0$ is the 1-dimensional subspace with the projector $\Pi_0$,
\begin{equation}
  W_0=\Pi_0 (V\otimes V),\quad \Pi_0=\frac{1}{4}\mathbb{Q}_{12}.
  \end{equation}
$W_s$ is the 9-dimensional subspace with the projector $\Pi_s$,
\begin{equation}
  W_s=\Pi_s (V\otimes V),\quad \Pi_s=\frac{1}{2}(\mathbb{I}+\mathbb{P}_{12})-\frac{1}{4}\mathbb{Q}_{12},
  \end{equation}
and $W_{\pm}$ are the 3-dimensional self-dual and anti-self-dual subspaces, respectively, with the projectors $\Pi_{\pm}$,
\begin{equation}
  W_{\pm}=\Pi_{\pm} (V\otimes V),\quad \Pi_{\pm}=\frac{1}{4}(\mathbb{I}-\mathbb{P}_{12})\pm \frac{1}{2}\mathbb{E}_{12},
  \end{equation}
where $\mathbb{E}$ is defined as
\begin{equation}
  \mathbb{E}_{12}|k,l\rangle=\frac{1}{2}\epsilon_{ijkl}|i,j\rangle.
  \end{equation}
Therefore, by Schur's lemma, we have
 \begin{equation}\label{eq:4site m12p12}
  \widehat{\mathbb{M}}_{12}\mathbb{P}_{12}=a \mathbb{I}+b \mathbb{P}_{12}+c \mathbb{Q}_{12}+d \mathbb{E}_{12},
  \end{equation}
with four coefficients $a,b,c,d$ to be determined. Next, we perform the following transformation for equation (\ref{eq:4site-central}),
\begin{equation}
  \begin{split}
&\Big[(S_L)_{12}A^t_3\Big] R_{13}(u)\bar{R}_{23}(u)\mathbb{M}_{12}\bar{R}_{13}(-u-2)R_{23}(-u-2)K_3(u)\Big[(S_R)_{12}A_3\Big]\\
=&\Big[(S_L)_{12}A^t_3\Big] K_3(u)\bar{R}_{13}(-u-2)R_{23}(-u-2)\mathbb{M}_{12}R_{13}(u)\bar{R}_{23}(u) \Big[(S_R)_{12}A_3\Big],
\end{split}
\end{equation}
and by defining
\begin{equation}
  \widehat{K}(u)=A^t K(u) A=\mathbb{I}+\frac{\widehat{K}^{(1)}}{u}+\frac{\widehat{K}^{(2)}}{u^2}+\cdots,
  \end{equation}
we find equation (\ref{eq:4site-central}) can be converted covariantly into
\begin{equation}\label{eq:4site-tr1}
\begin{split}
&R_{13}(u)\bar{R}_{23}(u)\widehat{\mathbb{M}}_{12}\bar{R}_{13}(-u-2)R_{23}(-u-2)\widehat{K}_3(u)\\
=&\widehat{K}_3(u) \bar{R}_{13}(-u-2)R_{23}(-u-2)\widehat{\mathbb{M}}_{12}R_{13}(u)\bar{R}_{23}(u).
\end{split}
\end{equation}
We now substitute $\widehat{\mathbb{M}}_{12}=a \mathbb{P}_{12}+b \mathbb{I}+c\mathbb{Q}_{12}- d \mathbb{E}_{12}$ from equation (\ref{eq:4site m12p12}) into equation (\ref{eq:4site-tr1}) and focus on the $u^{-2}$-order terms. For notational convenience, we define
\begin{eqnarray} \label{eq:4site nf1}
X_{123}=\mathbb{Q}_{23}-\mathbb{P}_{13}-2\mathbb{I},&\quad &X'_{123}=\mathbb{P}_{13} \mathbb{Q}_{23}-2\mathbb{P}_{13},\\ \label{eq:4site nf2}
Y_{123}=\mathbb{P}_{23}-\mathbb{Q}_{13}-2\mathbb{I},&\quad &Y'_{123}=\mathbb{Q}_{13} \mathbb{P}_{23}-2\mathbb{Q}_{13},
\end{eqnarray}
and then the $u^{-2}$-order terms can be expressed as
\begin{equation}\label{eq:4site-tr2}
\begin{split}
&\left[\widehat{K}^{(1)}_3, X_{123}\widehat{\mathbb{M}}_{12}+\widehat{\mathbb{M}}_{12}Y_{123}\right]\\
=&-X_{123}\widehat{\mathbb{M}}_{12}Y_{123}+Y_{123}\widehat{\mathbb{M}}_{12}X_{123}+\left[X'_{123}, \widehat{\mathbb{M}}_{12}\right]-\left[Y'_{123},\widehat{\mathbb{M}}_{12}\right].
\end{split}
\end{equation}
To extract useful information for coefficients $a,b,c,d$, we evaluate the matrix elements of the above equation between appropriate basis vectors on $V^{\otimes 3}$. There are many possible choices, and for concreteness, we choose the following ones:
\begin{itemize}
  \item $\langle 1,2,1|(\ref{eq:4site-tr2})|2,1,1 \rangle$ : $2a+b=0$;
  \item $\langle 1,1,1|(\ref{eq:4site-tr2})|1,1,1 \rangle$ : $c=0$;
  \item $\langle 1,2,1|(\ref{eq:4site-tr2})|1,4,3 \rangle$ : $d=0$;
\end{itemize}
Therefore, up to an overall scalar factor,
\begin{equation}
\widehat{\mathbb M}_{12}=\mathbb P_{12}-2\mathbb I,
\end{equation}
and the original $\mathbb{M}_{12}$ is given by
\begin{equation}\label{eq:4site m12K12}
  \begin{split}
 \mathbb{ M}_{12}=(A^{-t}\otimes A)(\mathbb{P}_{12}-2\mathbb{I})(A^{-1}\otimes A^t)=\mathbb{P}_{12}-2 K^{(0)}\otimes (K^{(0)})^{-1}
  \end{split}
  \end{equation}
Then we substitute $\widehat{\mathbb M}_{12}=\mathbb P_{12}-2\mathbb I$ back into the full equation (\ref{eq:4site-tr1}) and take the partial trace over the first space $V_1$ to obtain
\begin{equation}
  \begin{split}
  &(u+2)(4u^2+3u+1)\left(\mathbb{P}_{23}\widehat{K}_3(u)-\mathbb{P}_{23}\widehat{K}_2(u)\right)\\
  =& u(u+1)(4u+1)\left(\mathbb{P}_{23}\widehat{K}_2(u)\mathbb{Q}_{23}-\mathbb{Q}_{23}\widehat{K}_2(u)\mathbb{P}_{23}\right).
  \end{split}
  \end{equation}
Multiplying both sides by $\mathbb{P}_{23}$ on the right of the above equation, and then taking the partial trace over the second space $V_2$, we obtain
 \begin{equation}\label{eq:4site 2trace}
  \begin{split}
  (u+2)(4u^2+3u+1)\Big[\tr(\widehat{K}(u))\mathbb{I}-4\widehat{K}(u)\Big]=u(u+1)(4u+1)\Big[\widehat{K}(u)-\widehat{K}(u)^t\Big].
  \end{split}
  \end{equation}
The last step is to decompose $\widehat{K}(u)$ as
\begin{equation}
  \widehat{K}(u)=\frac{1}{4}\tr(\widehat{K}(u))\mathbb{I}+S(u)+A(u),
  \end{equation}
with $S^t=S$, $\operatorname{tr}S=0$, and $A^t=-A$, and from equation (\ref{eq:4site 2trace}) we find: $S(u)=A(u)=0$. Therefore we finally obtain, also up to a scalar factor,
\begin{equation}
\widehat{K}(u)=\mathbb{I},
\end{equation}
and the original $K(u)$ is
\begin{equation}
K(u)=A^{-t}\widehat{K}(u)A^{-1}=A^{-t}A^{-1}=K^{(0)}.
\end{equation}
We conclude that: in the $\mathfrak{so}(4)$ case, the $K(u)$ matrix may be chosen to be any invertible symmetric matrix, up to an overall scalar factor, and the corresponding $\mathbb{M}_{12}$ is constructed from this symmetric matrix by the equation (\ref{eq:4site m12K12}).

\subsubsection{$\mathfrak{sp}(4)$ case: $(K^{(0)})^t=-K^{(0)}$}
Any invertible antisymmetric $K^{(0)}$ can be transformed into a canonical symplectic form:
\begin{equation}
  A^t K^{(0)} A=J=
  \begin{pmatrix}
    0 & 1 & 0 & 0 \\
    -1 & 0 & 0 & 0 \\
    0 & 0 & 0 & 1 \\
    0 & 0 & -1 & 0
  \end{pmatrix}.
  \end{equation}
In the antisymmetric case, we have
\begin{equation}
  \begin{split}
  G_{ab}\otimes \mathbb{I}-\mathbb{I}\otimes G^t_{ab}=-K^{(0)}(E_{ab}+E_{ba})\otimes \mathbb{I}-\mathbb{I}\otimes (E_{ab}+E_{ba})K^{(0)}
  \end{split}
  \end{equation}
  where
\begin{equation}
  K^{(0)}(E_{ab}+E_{ba})=A^{-t} J A^{-1}(E_{ab}+E_{ba})=A^{-t}\Big[J A^{-1}(E_{ab}+E_{ba}) A^{-t}\Big] A^t,
  \end{equation}
therefore, using the formula given in appendix~\ref{appx:matrix}, we have
\begin{equation}
  \begin{split}
  S^{\prime-1}_{12}(G_{ab}\otimes \mathbb{I}-\mathbb{I}\otimes G^t_{ab})S^{\prime}_{12}=-(X_{ab}\otimes \mathbb{I}+\mathbb{I}\otimes X_{ab}),
   \end{split}
  \end{equation}
where
\begin{eqnarray}
  &&S^{\prime-1}_{12}=A^t\otimes A^tK^{(0)}=A^t\otimes J A^{-1},\\
  &&X_{ab}=A^t K^{(0)}(E_{ab}+E_{ba}) A^{-t}=J A^{-1}(E_{ab}+E_{ba})A^{-t}.
  \end{eqnarray}
Since
\begin{equation}
  X^t_{ab}J+JX_{ab}=0,
  \end{equation}
we recognize $X_{ab}$ as generators of the $\mathfrak{sp}(4)$ algebra, and then equation (\ref{eq:4site commutant}) indicates
\begin{equation}\label{eq:4site antisymN}
  \Big[\widehat{\mathbb{N}}_{12}\mathbb{P}_{12},X\otimes \mathbb{I}+\mathbb{I}\otimes X\Big]=0,\quad \forall X\in \mathfrak{sp}(4)
  \end{equation}
where
\begin{equation}\label{eq:4site n12}
  \widehat{\mathbb{N}}_{12}=(A^t\otimes JA^{-1})\mathbb{M}_{12}(AJ\otimes A^{-t})=(\mathbb{I}\otimes J){\widehat{\mathbb{M}}}_{12}(J\otimes \mathbb{I}).
\end{equation}
The irreducible decomposition of the diagonal $\mathfrak{sp}(4)$ action on $V\otimes V$ is:
\begin{equation}
  V\otimes V=\mathbb{C}|J\rangle \oplus \Lambda_0^2 V \oplus S^2V ,\quad \Leftrightarrow \quad \mathbf{4}\otimes \mathbf{4}=\mathbf{1}\oplus\mathbf{5}\oplus \mathbf{10},
  \end{equation}
where the 1-dimensional subspace $\mathbb{C}|J\rangle$ is projected out by the ``symplectic trace operator'' defined as
\begin{equation}\label{eq:4site Qj}
 \mathbb Q_J=(\mathbb I\otimes J^{-1})\mathbb Q_{12}(\mathbb I\otimes J).
\end{equation}
$\Lambda_0^2 V$ is the 5-dimensional antisymmetric subspace without symplectic trace,
\begin{equation}
\Lambda_0^2 V=\Pi_5(V\otimes V),\quad \Pi_5=\frac{1}{2}(\mathbb{I}-\mathbb{P}_{12})-\frac{1}{4}\mathbb{Q}_J,
\end{equation}
and $S^2V$ is the 10-dimensional symmetric subspace,
\begin{equation}
  S^2V=\Pi_{10}(V\otimes V),\quad \Pi_{10}=\frac{1}{2}(\mathbb{I}+\mathbb{P}_{12}).
  \end{equation}
Therefore, by Schur's lemma, equation (\ref{eq:4site antisymN}) means $\widehat{\mathbb{N}}_{12}\mathbb{P}_{12}$ is a linear combination of the above three projectors, or equivalently, we may write
\begin{equation}
  \widehat{\mathbb{N}}_{12}\mathbb{P}_{12}=a I+b \mathbb{P}_{12}+c(\mathbb{Q}_J)_{12},\quad \Rightarrow \quad \widehat{\mathbb{N}}_{12}=a\mathbb{P}_{12}+bI-c(\mathbb{Q}_J)_{12}.
  \end{equation}
Then equations (\ref{eq:4site n12}) and (\ref{eq:4site Qj}) give
\begin{equation}
\widehat{\mathbb M}_{12}=\alpha\mathbb P_{12}+\beta J_1J_2+\gamma\mathbb Q_{12}.
\end{equation}
The coefficients $\alpha,\beta$ and $\gamma$ are fixed by inserting the above $\widehat{\mathbb M}_{12}$ into equation (\ref{eq:4site-tr1}) and analyzing the $u^{-2}$-order term, which is given by
\begin{equation}\label{eq:4site coeff}
\begin{split}
&\widehat{K}^{'(1)}_3\left(Y_{123}\widehat{\mathbb{\mathbb{M}}}_{12}+\widehat{\mathbb{M}}_{12}X_{123}\right)-\left(X_{123}\widehat{\mathbb{M}}_{12}+\widehat{\mathbb{M}}_{12}Y_{123}\right)\widehat{K}^{'(1)}_3\\
=&J_3\Big[Y_{123}\widehat{\mathbb{\mathbb{M}}}_{12}X_{123}-Y'_{123}\widehat{\mathbb{\mathbb{M}}}_{12}-\widehat{\mathbb{\mathbb{M}}}_{12}X'_{123}\Big]-
\Big[X_{123}\widehat{\mathbb{\mathbb{M}}}_{12}Y_{123}-X'_{123}\widehat{\mathbb{\mathbb{M}}}_{12}-\widehat{\mathbb{\mathbb{M}}}_{12}Y'_{123}\Big]J_3,
\end{split}
\end{equation}
where
\begin{equation}
  \widehat{K}'(u)=A^t K(u) A=J+\frac{\widehat{K}^{'(1)}}{u}+\frac{\widehat{K}^{'(2)}}{u^2}+\cdots,
  \end{equation}
and $X_{123},X'_{123},Y_{123},Y'_{123}$ are the same as given in equations (\ref{eq:4site nf1}) and (\ref{eq:4site nf2}). By evaluating specific matrix elements of equation (\ref{eq:4site coeff}), we find:
\begin{itemize}
  \item $\langle 1,1,1|(\ref{eq:4site coeff})|1,1,2 \rangle$ : $\gamma=0$;
  \item $\langle 1,3,1|(\ref{eq:4site coeff})|3,1,2 \rangle$ : $\beta=2\alpha$;
\end{itemize}
We may therefore normalize
\begin{equation}
 \widehat{\mathbb M}_{12}=\mathbb P_{12}+2J_1J_2,
\end{equation}
and the original $\mathbb{M}_{12}$ is
\begin{equation}\label{eq:4site m12aK12}
  \begin{split}
 \mathbb{M}_{12}=&(A^{-t}\otimes A)\widehat{\mathbb{M}}_{12}(A^{-1}\otimes A^t)
 =(A^{-t}\otimes A)(\mathbb{P}_{12}+2J\otimes J)(A^{-1}\otimes A^t)\\
 =&\mathbb{P}_{12}-2K^{(0)}\otimes (K^{(0)})^{-1}.
  \end{split}
  \end{equation}
To determine $\widehat{K}'(u)$, we substitute the above $\widehat{\mathbb M}_{12}$ into the full equation (\ref{eq:4site-tr1}) and successively take partial traces over $V_1$ and $V_2$ to obtain
\begin{equation}\label{eq:4site anti}
\begin{split}
 &(2u^2+3u-4)\widehat{K}'(u))+(u+2)^2J\widehat{K}'(u))J-u(u+3)J\widehat{K}'(u)^tJ \\
 &+u(u+2)\operatorname{tr}\widehat{K}'(u))\,\mathbb I-(u+2)\operatorname{tr}\!\left[J\widehat{K}'(u))\right]J=0.
\end{split}
\end{equation}
We now decompose $\widehat{K}'(u)$ as
\begin{equation}
\widehat{K}'(u)=S(u)+A(u)-\frac{1}{4}\tr(J \widehat{K}'(u)) J,
\end{equation}
with $S^t(u)=S(u)$, $A^t(u)=-A(u)$ and $\tr(JA(u))=0$. The symmetric part of (\ref{eq:4site anti}) leads to the following two equations,
\begin{eqnarray}
\left\{
\begin{array}{lll}
(2u^2+3u-4)S(u)+(u+4)JS(u)J=0,\\
(2u^2+3u-4)JS(u)J+(u+4)S(u)=0.
\end{array}
\right.
\end{eqnarray}
For generic $u$, the solution is $S(u)=0$.  Similarly, the antisymmetric part of (\ref{eq:4site anti}) gives
\begin{eqnarray}
\left\{
\begin{array}{lll}
(2u^2+3u-4) A(u)+(2u^2+7u+4)JA(u)J=0,\\
(2u^2+3u-4) JA(u)J+(2u^2+7u+4)A(u)=0,
\end{array}
\right.
\end{eqnarray}
and the solution is $A(u)=0$. Consequently, $\widehat{K}'(u)$ is proportional to $J$, and can be chosen as $\widehat{K}'(u)=J$. The corresponding $K(u)$ is then
\begin{equation}
  K(u)=A^{-t}J A^{-1}=K^{(0)}.
  \end{equation}
Therefore, we see that, in the $\mathfrak{sp}(4)$ case, $K(u)$ can be any invertible antisymmetric matrix, and the corresponding $\mathbb{M}_{12}$ is constructed from this $K(u)$ by equation (\ref{eq:4site m12aK12}).

\subsubsection{$\mathfrak{sl}\,(4)$ case: $(K^{(0)})^t\neq \pm K^{(0)}$}

We now consider a generic invertible matrix $(K^{(0)})^t\neq\pm K^{(0)}$ and suppose that the corresponding matrices $G_{ab}$ generate the full
$\mathfrak{sl}(4)$ algebra. Equation (\ref{eq:4site commutant}) shows that $\mathbb{M}_{12}\mathbb{P}_{12}$ commutes with the diagonal action of $\mathfrak{sl}(4)$ on $V\otimes V^*$:
\begin{equation}
\Big[\mathbb{M}_{12}\mathbb{P}_{12}, X\otimes \mathbb{I}+\mathbb{I}\otimes X^*\Big]=0,\quad \forall X\in \mathfrak{sl}(4).
\end{equation}
Since
\begin{equation}
 \boldsymbol{4}\otimes\boldsymbol{4}^*=\boldsymbol{1}\oplus\boldsymbol{15},
\end{equation}
Schur's lemma restricts the operator to
\begin{equation}
 \mathbb M_{12}\mathbb P_{12}
 =a\mathbb I+b\mathbb Q_{12},
 \qquad\text{or equivalently}\qquad
 \mathbb M_{12}
 =\mu\mathbb P_{12}+\nu\mathbb Q_{12}.
\label{eq:four-site-generic-ansatz}
\end{equation}
The $u^{-2}$-order of the equation (\ref{eq:4site-central}) then gives $\mu=0$ and $\nu=0$. Thus the generic $\mathfrak{sl}(4)$ sector
contains no nonzero four-site block.

In summary, we find that for the two nontrivial $\mathfrak{so}(4)$ and $\mathfrak{sp}(4)$ branches, the chiral 4-site integrable product states are generated by
\begin{equation}
 K(u)=K,\qquad K^t=\pm K,\qquad
 \mathbb M_{12}
 =\mathbb P_{12}-2K_1K_2^{-1}
\end{equation}
with $\det K\neq0$.  In components, the corresponding local 4-site block is
\begin{equation}
 |\phi\rangle_{1234}=\left[\delta^i_l\delta^j_k-2K^i_k(K^{-1})^j_l\right]|i,j,k,l\rangle_{1234}.
\end{equation}
We note that these results coincide with the 4-site states obtained from the fusion construction of reflection matrices \cite{Liu:2026abjmre}, and have also been checked using a \textit{Mathematica} program.

\subsection{Achiral 4-site states}
The basic state-operator equation for an achiral block $|\phi\rangle$ is
\begin{equation}
\begin{split}
&R_{01}(u)\bar{R}_{02}(u)R_{03}(u)\bar{R}_{04}(u)|\phi\rangle_{1234}K_0(u)\\
=&K_0(u)R_{01}(-u-2)\bar{R}_{02}(-u-2)R_{03}(-u-2)\bar{R}_{04}(-u-2)|\phi\rangle_{1234},
\end{split}
\end{equation}
which can be rewritten as the following pure operator equation:
\begin{equation}
\begin{split}
&R_{13}(u)\bar{R}_{23}(u)\mathbb{M}_{12}\bar{R}_{13}(-u-2)R_{23}(-u-2)K_3(u)\\
=&K_3(u){R}_{13}(-u-2)\bar{R}_{23}(-u-2)\mathbb{M}_{12}\bar{R}_{13}(u){R}_{23}(u).
\end{split}
\end{equation}
Using $M^{ij}_{kl}:=\langle i,j|\mathbb{M}_{12}|k,l\rangle$, and denoting $K:=K(0)$, the component form of the above equation is
\begin{equation}\label{eq:4site extra f1}
\begin{split}
&\delta_{xy}\delta_{ab}M^{zr}_{rd}K^d_c-2\delta_{xy}K^a_c M^{zr}_{rb}-2\delta_{ab}M^{zy}_{xd}K^d_c+4K^a_cM^{zy}_{xb}\\
=&\delta_{xy}\delta_{ab}K^z_pM^{pr}_{rc}-2\delta_{xy}K^z_pM^{pb}_{ac}-2\delta_{ab}K^z_yM^{xr}_{rc}+4K^z_yM^{xb}_{ac}.
\end{split}
\end{equation}
For $x\neq y$, $a\neq b$, the above equation reduces to
\begin{equation}\label{eq:4site f2}
K^a_cM^{zy}_{xb}=K^z_yM^{xb}_{ac}.
\end{equation}
Since $K(u)$ is assumed to be invertible for generic $u$ including $u=0$,
\begin{equation}\label{eq:4site f1}
M^{xb}_{ac}=\sum_z(K^{-1})^y_zM^{zy}_{xb}K^a_c:=L^x_b K^a_c,\quad  a\neq b.
\end{equation}
Next we prove $L^x_b\propto K^x_b$. Similarly to (\ref{eq:4site f1}), we also have
\begin{equation}
M^{zy}_{xb}=L^z_y K^x_b,
\end{equation}
thus equation (\ref{eq:4site f2}) is
\begin{equation}
K^a_c L^z_y K^x_b=K^z_yL^x_b K^a_c,\quad \Rightarrow \quad L^z_y K^x_b=K^z_yL^x_b,\quad x\neq y,
\end{equation}
which implies
\begin{equation}
L^x_b=\lambda K^x_b,
\end{equation}
and thus
\begin{equation}\label {eq:4site f3}
M^{xb}_{ac}=\lambda K^x_b K^a_c,\quad a\neq b.
\end{equation}
We now return to equation (\ref{eq:4site extra f1}). For $a=b$ and $x\neq y$, using relation (\ref{eq:4site f3}) and cancelling the common factor $K^z_y$, we obtain
\begin{equation}\label{eq:4site f4}
-2\lambda (K^2)^x_c+4\lambda K^a_c K^x_a=-2 M^{xr}_{rc}+4M^{xa}_{ac},
\end{equation}
where no summation over the index $a$ is understood. Since this equation holds for each value of $a$, taking the sum over it gives
\begin{equation}
M^{xr}_{rc}=\lambda(K^2)^x_c.
\end{equation}
Therefore, equation (\ref{eq:4site f4}) reduces to
\begin{equation}\label{eq:4site f5}
M^{xa}_{ac}=\lambda K^x_a K^a_c.
\end{equation}
Combining the results (\ref{eq:4site f3}) and (\ref{eq:4site f5}), we find in general,
\begin{equation}
M^{ij}_{kl}=\lambda K^i_j K^k_l,\quad \forall i,j,k,l.
\end{equation}
Thus the basic achiral 4-site block is
\begin{equation}
\begin{split}
|\phi\rangle_{1234}=M^{ij}_{kl}|i,j,k,l\rangle=\lambda K^i_j K^k_l|i,j,k,l\rangle=\lambda |m\rangle_{12}\otimes |m\rangle_{34},
\end{split}
\end{equation}
where
\begin{equation}
|m\rangle_{12}=K^i_j|i,j\rangle,
\end{equation}
showing that every achiral 4-site block is essentially the product of two achiral 2-site states.

\section{Conclusions}
In this paper we have studied chiral and achiral integrable boundary states in
the alternating ABJM spin chain by solving integrable pairs
$(|\Psi\rangle,K(u))$ satisfying the $KT$-relation.  A central feature of our
approach is that, except for the 3-site cases, we do not impose any reflection
equation on $K(u)$ in advance.  Instead, $K(u)$ and the corresponding boundary
block are determined simultaneously from the mixed state-operator constraints
following from the $KT$-relation.

We focused on boundary states with $n$-site translational invariance for
$n\leq 4$.  For each block size, the state is constructed from an elementary
$n$-site block, and the $KT$-relation gives a fundamental mixed equation
involving this block and $K(u)$.  For odd block sizes, namely the 1-site and
3-site cases, this equation can be transformed into a state equation on the
quantum space.  For even block sizes, namely the 2-site and 4-site cases, it can
instead be reduced to an operator equation.  In the most general setting, the
$n$-site block and $K(u)$ may carry additional internal degrees of freedom and
therefore become operator-valued objects acting on an internal space.  We
analyzed this general operator-valued solution in the 1-site case.  For the
remaining cases, we restricted our attention to $c$-number solutions.

Our results are summarized in Table~\ref{tab:conclusion-summary}.

\begin{table}[t]
\centering
\renewcommand{\arraystretch}{1.25}
\begin{tabular}{c|c|p{0.62\textwidth}}
\hline
Block size & Type & Result \\
\hline
1-site
& chiral
& Nontrivial operator-valued solutions exist.  The 1-site block is generated
by Clifford algebra generators, and the corresponding $K(u)$ is also
constructed from Clifford generators.  In particular, $K(u)$ is invertible. \\
\hline
1-site
& achiral
& No nontrivial integrable boundary state exists. \\
\hline
2-site
& chiral
& No nontrivial $c$-number solution exists. \\
\hline
2-site
& achiral
& Nontrivial $c$-number integrable pairs $(M,K(u))$ exist.  Here $M$ determines
the 2-site elementary block, and $K(u)$ is fixed by $M$.  The resulting
$K(u)$ necessarily satisfies the SP-type reflection equation. \\
\hline
3-site
& chiral
& When $K(u)$ is chosen from the SNP-type reflection-equation solutions, no
nontrivial $c$-number integrable boundary state exists. \\
\hline
3-site
& achiral
& When $K(u)$ is chosen from the SP-type reflection-equation solutions, no
nontrivial $c$-number integrable boundary state exists. \\
\hline
4-site
& chiral
& The solutions are classified according to the constant term in the Laurent
expansion of $K(u)$.  In the $\mathfrak{so}(4)$ and $\mathfrak{sp}(4)$ cases, $K(u)$ is an
invertible spectral-parameter-independent constant matrix, symmetric in the
$\mathfrak{so}(4)$ case and antisymmetric in the $\mathfrak{sp}(4)$ case.  In the $\mathfrak{sl}(4)$ case, only
the zero solution exists. \\
\hline
4-site
& achiral
& The 4-site solutions essentially factorize into tensor products of two
2-site solutions.  Hence this case reduces to the 2-site achiral integrable
states. \\
\hline
\end{tabular}
\caption{Summary of the chiral and achiral integrable boundary states with
$n$-site translational invariance for $n\leq 4$.  Except for the 1-site chiral
case, where operator-valued solutions are considered, the classification refers
to $c$-number solutions.}
\label{tab:conclusion-summary}
\end{table}

This classification suggests several directions for future work.  First, it would be interesting to study operator-valued integrable pairs for
$n>1$. Even for $n=1$, it remains interesting to look for operator-valued solutions beyond the Clifford-algebraic realization constructed in this work.

Second, the 1-site and 3-site analyses suggest that nontrivial $c$-number integrable pairs may be absent for general $(2k+1)$-site translationally
invariant blocks.  It would be useful either to prove this conjecture or to find a counterexample at higher odd block size.

Third, it would be important to compute the overlaps between the $n$-site integrable states constructed here and Bethe eigenstates of the ABJM spin
chain.  Such overlap formulas would be directly relevant for one-point and higher-point functions in defect setups of the ABJM theory.

Finally, it would be interesting to understand whether the $n$-site boundary states found here admit an effective field-theory interpretation, for example
in terms of specific boundary or defect conditions in the dual field theory. Such a relation would clarify which boundary or defect observables in the dual
field theory are captured by the integrable boundary states constructed in this paper.

\section*{Acknowledgments}
Nan Bai would like to thank Jun-Bao Wu for carefully reading the manuscript and for helpful discussions. This work was supported by the National Natural Science Foundation of China (Grant No. 12165002).

\begin{appendix}
\section{Solutions of SP-type reflection equation}\label{app:SP}
In this appendix we rederive the solutions of the SP-type reflection equation given in \cite{Arnaudon:2004sd} using a slightly different approach. The SP-type reflection equation takes the form
\begin{equation}\label{apd1:f1}
R_{12}(u-v)K_1(u)R_{12}(u+v)K_2(v)=K_2(v)R_{12}(u+v)K_1(u)R_{12}(u-v).
\end{equation}
Substituting the explicit $R$-matrix into this equation, we obtain
\begin{equation}
\begin{split}
\frac{1}{u^2-v^2}\left[K_2(u)K_2(v)-K_2(v)K_2(u)\right]=\frac{1}{u+v}\left[-K_1(u)\mathbb{P}_{12}K_2(v)+K_2(v)\mathbb{P}_{12}K_1(u)\right]\\
+\frac{1}{u-v}\left[-\mathbb{P}_{12}K_1(u)K_2(v)+K_2(v)K_1(u)\mathbb{P}_{12}\right].
\end{split}
\end{equation}
Taking the partial trace over $V_2$, we obtain
\begin{equation}\label{apd1:f2}
\Big[K(u),K(v)\Big]=0.
\end{equation}
Therefore, the SP-type reflection equation is equivalent to
\begin{equation}\label{apd1:f4}
\frac{1}{u+v}\Big[K_2(v)K_2(u)-K_1(u)K_1(v)\Big]+\frac{1}{u-v}\Big[K_2(v)K_1(u)-K_2(u)K_1(v)\Big]=0.
\end{equation}
Let $v=-u$ in the above equation, we find
\begin{equation}
K_2(u)K_2(-u)=K_1(u)K_1(-u),
\end{equation}
showing that,
\begin{equation}\label{apd1:f3}
K(u)K(-u)=f(u) \mathbb{I},
\end{equation}
which is a well-known result of the reflection algebras \cite{Molev 2002}.

Suppose that $K(u)$ has the Laurent expansion
\begin{equation}
  K(u)=G+\frac{K^{(1)}}{u}+\frac{K^{(2)}}{u^2}+\cdots,\quad G\neq 0,
\end{equation}
and the equation (\ref{apd1:f2}) gives
\begin{equation}
  \left[K(u),G\right]=\left[K(u),K(\infty)\right]=0,
  \end{equation}
indicating that the eigenspaces of $G$ are $K(u)$-invariant. We next classify the solutions by the leading matrix $G$. By taking $v\rightarrow\infty$, the equation (\ref{apd1:f4}) reduces to
\begin{equation}\label{apd1:f5}
 K_2(u)G_1-G_2K_1(u)-K_1(u)G_1+G_2K_2(u)=0.
 \end{equation}
Taking $u\rightarrow\infty$, the relation (\ref{apd1:f3}) gives
\begin{equation}
G^2=a \mathbb{I}.
\end{equation}
We now distinguish two cases:

\noindent $\bullet$ $a\neq 0$: In this case $G^2$ can be chosen as $G^2=\mathbb{I}$ by the rescaling $K(u)\rightarrow cK(u)$.  Since the $R$-matrix has the property that, for any invertible matrix $A$,
\begin{equation}
  (A\otimes A)R_{12}(u)(A^{-1}\otimes A^{-1})=R_{12}(u),
\end{equation}
we may choose an appropriate $A$ and use the similarity transformation $K(u)\rightarrow A K(u) A^{-1}$ to bring $G$ to the standard form:
\begin{equation}
  G={\rm{diag}}(\mathbb{I}_{m\times m},-\mathbb{I}_{n\times n}).
\end{equation}
In this eigenbasis of $G$, $K(u)$ takes the block-diagonal form:
\begin{equation}
  K(u)=
  \begin{pmatrix}
  K^{+}(u)&0   \\
    0&K^-(u)
  \end{pmatrix}.
  \end{equation}
We denote $V^{\pm}=\{v\in V|Gv=\pm v\}$. Then in the subspace $V^+\otimes V^+$, the equation (\ref{apd1:f5}) reduces to
\begin{equation}
K^+_2(u)-K^+_1(u)-K^+_1(u)+K^+_2(u)=0,\quad \Rightarrow \quad K^+_1(u)=K^+_2(u),
\end{equation}
showing that $K^+(u)$ is a scalar matrix: $K^+(u)=k^+(u)\mathbb{I}_{m\times m}$. In the subspace $V^-\otimes V^-$, the equation (\ref{apd1:f5}) reduces to
\begin{equation}
-K^-_2(u)+K^-_1(u)+K^-_1(u)-K^-_2(u)=0,\quad \Rightarrow \quad K^-_1(u)=K^-_2(u),
\end{equation}
showing that the lower block is also scalar: $K^-(u)=k^-(u)\mathbb{I}_{n\times n}$. To fix $k^+(u)$ and $k^-(u)$, we turn to the full reflection equation (\ref{apd1:f4}) and in the mixed sector $V^+\otimes V^-$, we have
\begin{equation}
 (u+v)\left(k^+(v)k^-(u)-k^+(u)k^-(v)\right)+(u-v)\left(k^+(u)k^+(v)-k^-(u)k^-(v)\right)=0.
 \end{equation}
Letting $r(u)=k^+(u)/k^-(u)$, the above equation can be written in the form of separated variables as
\begin{equation}
  \frac{(r(v)+1)v}{1-r(v)}=\frac{(r(u)+1)u}{1-r(u)},
  \end{equation}
which implies that the result is a constant independent of $u$,
\begin{equation}
  \frac{(r(u)+1)u}{1-r(u)}=\lambda \quad \Rightarrow \quad r(u)=\frac{k^+(u)}{k^-(u)}=\frac{\lambda-u}{\lambda+u}.
  \end{equation}
Therefore, up to a scalar function of $u$, $K(u)$ takes the following form:
\begin{equation}
  K(u)=
  \begin{pmatrix}
    (\lambda-u)\mathbb{I}_m & 0 \\
    0 & (\lambda+u)\mathbb{I}_n
  \end{pmatrix}=\lambda \mathbb{I}-u G.
  \end{equation}
\noindent $\bullet$ $a=0$: Then $G$ is a second-order nilpotent matrix whose eigenvalues are zero, and thus,
\begin{equation}
\tr G=0.
\end{equation}
We now take the partial trace over $V_2$ of the equation (\ref{apd1:f5}) to obtain
\begin{equation}\label{apd1:f6}
(\tr K(u))G-(\tr G)K(u)-4 K(u) G+\tr(GK(u))\mathbb{I}=0.
\end{equation}
Multiplying by $G$ on the right and using the result $\tr G=0$, we find,
\begin{equation}
\tr(GK(u)) G=0,\quad \Rightarrow \quad \tr(GK(u))=0.
\end{equation}
The equation (\ref{apd1:f6}) therefore becomes
\begin{equation}
(\tr K(u))G=4 K(u) G\quad \Rightarrow \quad K(u)G=\frac{\tr K(u)}{4}G=\alpha(u) G.
\end{equation}
Substituting this result back into equation (\ref{apd1:f5}), we obtain
\begin{equation}
G_1(K_2(u)-\alpha(u) \mathbb{I})=(K_1(u)-\alpha(u) \mathbb{I})G_2,
\end{equation}
which implies
\begin{equation}
K(u)=\alpha(u)\mathbb{I}+\beta(u)G.
\end{equation}
Substituting this form of $K(u)$ into the full reflection equation (\ref{apd1:f4}), we have
\begin{equation}
\left[\frac{\alpha(u)\beta(v)+\alpha(v)\beta(u)}{u+v}-\frac{\alpha(v)\beta(u)-\alpha(u)\beta(v)}{u-v}\right](G_2-G_1)=0,
\end{equation}
which leads to
\begin{equation}
\frac{u\alpha(u)}{\beta(u)}=\frac{v\alpha(v)}{\beta(v)}=\lambda.
\end{equation}
Thus, up to a scalar function, we obtain
\begin{equation}
K(u)=\lambda \mathbb{I}+u G.
\end{equation}
We summarize the above analysis as follows. Up to an overall scalar factor and a similarity transformation, the solutions of the SP-type reflection equation take the generic form
\begin{equation}
K(u)=c\mathbb{I}+u G,
\end{equation}
where $G$ can be chosen as either of the following two forms:
\begin{itemize}
  \item [(a)] $G^2=\mathbb{I}$. In the case ${\rm{dim}}V=4$, $G$ has the canonical form
  \begin{equation}
  G={\rm{diag}}(\mathbb{I}_{n},-\mathbb{I}_{4-n}),\quad n\leq 4.
  \end{equation}
  \item [(b)] $G^2=0$. The canonical form of the second-order nilpotent block is
   \begin{equation}
J_2(0)=\begin{pmatrix}
         0 & 1 \\
         0 & 0
       \end{pmatrix}.
\end{equation}
Thus, in the case ${\rm{dim}}V=4$, $G$ has two choices:
\begin{equation}
G\sim J_2(0)\oplus J_2(0)\quad {\mbox{or}}\quad G\sim J_2(0)\oplus 0\oplus 0.
\end{equation}
\end{itemize}

\section{Solutions of  SNP-type reflection equation}\label{app:SNP}
We now study the solutions of the SNP-type reflection equation. The invertible solution $K(u)$ was discussed in \cite{Arnaudon:2004sd}.  Here we give a complete investigation of the solutions including non-invertible ones.

The SNP-type reflection equation is
\begin{equation}
R_{12}(u-v) K_1(u) \bar{R}_{12}(u+v) K_2(v)=K_2(v) \bar{R}_{12}(u+v) K_1(u) R_{12}(u-v).
\end{equation}
Substituting the explicit expressions for $R$- and $\bar{R}$-matrices into the above equation, we obtain
\begin{equation}\label{apd2:f3}
\begin{split}
&(u-v)\Big[K_1(u) \mathbb{Q}_{12} K_2(v)-K_2(v) \mathbb{Q}_{12} K_1(u)\Big]\\
=&(u+v+2)\Big[\mathbb{P}_{12} K_1(u) K_2(v)-\mathbb{P}_{12} K_1(v) K_2(u)\Big]\\
&-\Big[\mathbb{P}_{12} K_1(u) \mathbb{Q}_{12} K_2(v)-\mathbb{P}_{12} K_1(v) \mathbb{Q}_{12} K_2(u)\Big].
\end{split}
\end{equation}
The symmetric part under the exchange of $u$ and $v$ can be extracted as
\begin{equation}
K_1(u) \mathbb{Q}_{12} K_2(v)-K_2(v) \mathbb{Q}_{12} K_1(u)=K_1(v) \mathbb{Q}_{12} K_2(u)-K_2(u) \mathbb{Q}_{12} K_1(v).
\end{equation}
Performing the partial transpose on $V_2$ and then multiplying $\mathbb{P}_{12}$ to the right, we obtain
\begin{equation}\label{apd2:f1}
K_1(u) K^t_2(v) - K^t_1(v) K_2(u)=K_1(v) K^t_2(u) - K^t_1(u) K_2(v).
\end{equation}
Taking the transpose in $V_2$ again, we have
\begin{equation}\label{apd2:f2}
K_1(u) K_2(v) - K^t_1(v) K^t_2(u)=K_1(v) K_2(u) - K^t_1(u) K^t_2(v).
\end{equation}
Equations (\ref{apd2:f1}) and (\ref{apd2:f2}) indicate that $K(u)$ has the general form
\begin{equation}
K(u)=p(u) S+q(u) A,
\end{equation}
where $S$ and $A$ are constant symmetric and antisymmetric matrices independent of $u$, respectively: $S^t=S,\,A^t=-A$. Substituting this form of $K(u)$ back into the full reflection equation (\ref{apd2:f3}), we obtain
\begin{equation}\label{apd2:f4}
C(u,v)(A_1 \mathbb{Q}_{12} S_2-S_1 \mathbb{Q}_{12} A_2)+D(u,v)(S_1 A_2-A_1 S_2)=0,
\end{equation}
where
\begin{eqnarray}
C(u,v)&=&(u-v+1)p(u) q(v)+(u-v-1)p(v) q(u),\\
D(u,v)&=&(u+v+2)\Big[p(u) q(v)-p(v) q(u)\Big].
\end{eqnarray}
The positions of $C(u,v)$ and $D(u,v)$ can also be interchanged.  This is achieved by first taking the transpose in $V_2$, then multiplying by $\mathbb{P}_{12}$ on the right, and finally taking the transpose in $V_2$ again.  In this way we obtain
\begin{equation}\label{apd2:f5}
D(u,v)(A_1 \mathbb{Q}_{12} S_2-S_1 \mathbb{Q}_{12} A_2)+C(u,v)(S_1 A_2-A_1 S_2)=0.
\end{equation}
We now define
\begin{eqnarray}\label{apd2:f6}
W_{12}&=&A_1 \mathbb{Q}_{12} S_2-S_1 \mathbb{Q}_{12} A_2,\\ \label{apd2:f7}
Z_{12}&=&S_1 A_2-A_1 S_2,
\end{eqnarray}
then equations (\ref{apd2:f4}) and (\ref{apd2:f5}) are equivalent to the following two constraints:
\begin{eqnarray}
&&\Big[C(u,v)-D(u,v)\Big]\left(W_{12}-Z_{12}\right)=0,\\
&&\Big[C(u,v)+D(u,v)\Big]\left(W_{12}+Z_{12}\right)=0.
\end{eqnarray}
Therefore, to satisfy the above two equations, we distinguish the following four cases.

\noindent $\bullet$ $C(u,v)-D(u,v)=0,\,C(u,v)+D(u,v)=0$: The solution is $p(u)=0$ or $q(u)=0$, and then $K(u)$ is constant symmetric or antisymmetric matrix, up to an overall scalar factor.

\noindent $\bullet$ $C(u,v)-D(u,v)=0,\,W_{12}+Z_{12}=0$: The first condition $C(u,v)=D(u,v)$ gives
\begin{equation}
   \frac{p(u)}{(2u+1)q(u)}=\frac{p(v)}{(2v+1)q(v)},
\end{equation}
thus
\begin{equation}
    \frac{p(u)}{(2u+1)q(u)}=\lambda,
\end{equation}
where $\lambda$ is a constant, and then $K(u)$ reduces to the form
\begin{equation}
   K(u)=(2u+1)S+A.
\end{equation}
The second condition $W_{12}+Z_{12}=0$ is
\begin{equation}
   A_1 \mathbb{Q}_{12} S_2-S_1 \mathbb{Q}_{12} A_2+S_1 A_2-A_1 S_2=0.
   \end{equation}
The defining equation (\ref{apd2:f6}) of $W_{12}$ indicates
\begin{equation}
   {\rm{\mathbf{rank}}} W_{12}\leq {\rm{\mathbf{rank}}}(A_1 \mathbb{Q}_{12} S_2)+{\rm{\mathbf{rank}}}(S_1 \mathbb{Q}_{12} A_2)\leq 2\,{\rm{\mathbf{rank}}} (\mathbb{Q}_{12})=2.
   \end{equation}
It then follows from the defining equation (\ref{apd2:f7}) of $Z_{12}$ and the relation $Z_{12}=-W_{12}$ that
\begin{equation}
  {\rm{\mathbf{rank}}} Z_{12}={\rm{\mathbf{rank}}}(S\otimes A-A\otimes S)\leq 2.
   \end{equation}
The rank of the antisymmetric matrix $A$ can be 0, 2 and 4, so we discuss these three cases separately:
\begin{itemize}
  \item [(a)] ${\rm{\mathbf{rank}}}(A)=4$: In this case, $A$ is invertible, and hence $Z_{12}$ can be written as
  \begin{equation}
     Z_{12}=(A\otimes A)\left(A^{-1} S\otimes \mathbb{I}-\mathbb{I}\otimes A^{-1} S\right),
  \end{equation}
  and thus
  \begin{equation}\label{apd2:f8}
   {\rm{\mathbf{rank}}} Z_{12}={\rm{\mathbf{rank}}} \left(A^{-1} S\otimes \mathbb{I}-\mathbb{I}\otimes A^{-1} S\right)\leq 2.
   \end{equation}
   If $S\neq 0$, then there exists $x\in V$ such that $x$ and $A^{-1}S\,x$ are linearly independent.  For the following four vectors on $V\otimes V$:
   \begin{equation}
   x\otimes e_j,\quad j=1,\cdots,4,
   \end{equation}
   after the action of $A^{-1} S\otimes \mathbb{I}-\mathbb{I}\otimes A^{-1} S$ $\in{\rm{End}}(V\otimes V)$,
   \begin{equation}
   (A^{-1} S\otimes \mathbb{I}-\mathbb{I}\otimes A^{-1} S) (x\otimes e_j)=A^{-1} Sx\otimes e_j-x\otimes A^{-1} S e_j,
   \end{equation}
   their images are still linearly independent, indicating that:
   \begin{equation}
   {\rm{\mathbf{rank}}} Z_{12}\geq 4,
   \end{equation}
   which contradicts the inequality (\ref{apd2:f8}). Thus $S$ must vanish and then $K(u)= A$.
  \item [(b)] ${\rm{\mathbf{rank}}} (A)=0$: Thus $A=0$ and $K(u)=S$.
  \item [(c)] ${\rm{\mathbf{rank}}} (A)=2$: In this case we first note that
  \begin{equation}
     {\rm{dim}\,\rm{Im}} A=2.
     \end{equation}
  For $\forall x\in {\rm{Ker}} A,\,\forall y\in V$,
  \begin{eqnarray}
     Z_{12}(x\otimes y)&=&Sx\otimes Ay,\\
     Z_{12}(y\otimes x)&=&-Ay\otimes Sx.
     \end{eqnarray}
  If $Sx\neq 0$, the above results imply
  \begin{equation}
     {\rm{\mathbf{rank}}}Z_{12}={\rm{dim}\,\rm{Im}}Z_{12}\geq 3,
     \end{equation}
     which again contradicts the inequality (\ref{apd2:f8}). Thus $Sx=0$, which implies
     \begin{equation}\label{apd2:f9}
     {\rm{Ker}} A\subseteq {\rm{Ker}} S.
     \end{equation}
     Moreover,
     \begin{equation}\label{apd2:f10}
     {\rm{dim}\, \rm{Ker}}A=4-{\rm{dim}\, \rm{Im}}A=2.
     \end{equation}
     Equations (\ref{apd2:f9}) and (\ref{apd2:f10}) restrict $A$ and $S$ to the forms:
     \begin{equation}\label{apd2:f11}
\setlength{\arraycolsep}{7pt}
\renewcommand{\arraystretch}{1.05}
A=
\begin{pmatrix}
0&a&0&0\\
-a&0&0&0\\
0&0&0&0\\
0&0&0&0
\end{pmatrix},\quad
     S=\begin{pmatrix}
     s_1&s_2&0&0\\
     s_2&s_3&0&0\\
     0&0&0&0\\
     0&0&0&0
     \end{pmatrix},
\end{equation}
     where $a\neq 0$. In this basis,  $W_{12}$ and $Z_{12}$ take the following restricted forms,
     \begin{equation}\label{apd2:f12}
     W_{12}=a\begin{pmatrix}
               0 & -s_1 & s_1 & 0 \\
               s_1 & 0 & 2s_2 & s_3 \\
              -s_1 & -2s_2 & 0 & -s_3 \\
               0 & -s_3 & s_3 & 0
             \end{pmatrix},\quad
          Z_{12}=a\begin{pmatrix}
               0 & s_1 & -s_1 & 0 \\
               -s_1 & 0 & -2s_2 & -s_3 \\
              s_1 & 2s_2 & 0 & s_3 \\
               0 & s_3 & -s_3 & 0
             \end{pmatrix}.
     \end{equation}
     We find the constraint $W_{12}+Z_{12}=0$ is satisfied automatically. Therefore we obtain a nontrivial solution of the SNP-type reflection equation:
     $K(u)=(2u+1)S+A$ with $S$ and $A$ given in (\ref{apd2:f11}). This solution is, however, non-invertible.
     \end{itemize}
     \noindent $\bullet$ $C(u,v)+D(u,v)=0,\,W_{12}-Z_{12}=0$: The first condition $C(u,v)+D(u,v)=0$ gives
     \begin{equation}
   \frac{(2u+3)p(u)}{q(u)}=\frac{(2v+3)p(v)}{q(v)},
   \end{equation}
   which implies that $K(u)$ should take the form
   \begin{equation}
   K(u)=S+(2u+3)A.
   \end{equation}
   The second condition $W_{12}-Z_{12}=0$ is
   \begin{equation}
   A_1 \mathbb{Q}_{12} S_2-S_1 \mathbb{Q}_{12} A_2=S_1 A_2-A_1 S_2.
   \end{equation}
   We use an analysis similar to that in case (c) above. In the ${\rm{\mathbf{rank}}} (A)=4$ and ${\rm{\mathbf{rank}}} (A)=0$ branches, we obtain $K(u)=A$ and $K(u)=S$, respectively. For the branch  ${\rm{\mathbf{rank}}} (A)=2$, with the concrete forms of $W_{12}$ and $Z_{12}$ given in (\ref{apd2:f12}), the relation $W_{12}=Z_{12}$ indicates $S=0$.  Thus $K(u)$ turns out to be an antisymmetric matrix: $K(u)=A$.
   
  \noindent $\bullet$ $W_{12}=0,\,Z_{12}=0$: $Z_{12}=0$ gives:
   \begin{equation}
   S_1 A_2=A_1 S_2\quad \Rightarrow \quad S_1 A^t_2=A_1 S^t_2\quad \Rightarrow \quad -S_1 A_2=A_1 S_2,
   \end{equation}
   hence
   \begin{equation}
   S_1 A_2=0\quad \Rightarrow \quad S=0,\,\mbox{or}\,\,A=0,
   \end{equation}
   so $K(u)=A$ or $K(u)=S$.

\section{A matrix equality}\label{appx:matrix}
We prove a simple matrix identity that is used in the main text. For arbitrary matrices $A$ and $B$, suppose $A$ is invertible, then
\begin{equation}
BA\equiv A^{-1}(AB)A,
\end{equation}
hence
\begin{equation}
\begin{split}
AB\otimes \mathbb{I}+\mathbb{I}\otimes BA
=(\mathbb{I}\otimes A^{-1})(AB\otimes \mathbb{I}+\mathbb{I}\otimes AB)(\mathbb{I}\otimes A).
\end{split}
\end{equation}
For an arbitrary matrix $E$ and an arbitrary invertible matrix $C$, since $E\equiv CC^{-1}ECC^{-1}$, we have
\begin{eqnarray}
E\otimes\mathbb{I}&=&(C\otimes C)(C^{-1}EC\otimes \mathbb{I})(C^{-1}\otimes C^{-1}),\\
\mathbb{I}\otimes E&=&(C\otimes C)(\mathbb{I}\otimes C^{-1}EC)(C^{-1}\otimes C^{-1}).
\end{eqnarray}
The above identities lead to the following decomposition:
\begin{equation}
\begin{split}
&AB\otimes \mathbb{I}+\mathbb{I}\otimes BA\\
=&(\mathbb{I}\otimes A^{-1})(C\otimes C)(C^{-1}ABC\otimes \mathbb{I}+\mathbb{I}\otimes C^{-1}ABC)(C^{-1}\otimes C^{-1})(\mathbb{I}\otimes A).
\end{split}
\end{equation}
Writing
\begin{equation}
S^{-1}_{12}=C\otimes A^{-1}C,\quad F=C^{-1}ABC,
\end{equation}
then
\begin{equation}
AB\otimes \mathbb{I}+\mathbb{I}\otimes BA\equiv S^{-1}_{12}(F\otimes \mathbb{I}+\mathbb{I}\otimes F) S_{12},
\end{equation}
which shows that $AB\otimes \mathbb{I}+\mathbb{I}\otimes BA$ is similar to the diagonal action of $F$ on $V\otimes V$.

\end{appendix}


\begin{thebibliography}{99}

\bibitem{Ghoshal:1993tm}
S.~Ghoshal and A.~B.~Zamolodchikov,
\textit{Boundary S matrix and boundary state in two-dimensional integrable quantum field theory},
Int. J. Mod. Phys. A \textbf{9} (1994), 3841-3886
[erratum: Int. J. Mod. Phys. A \textbf{9} (1994), 4353]
%doi:10.1142/S0217751X94001552
[arXiv:hep-th/9306002].

\bibitem{Piroli:2017sei}
L.~Piroli, B.~Pozsgay and E.~Vernier,
\textit{What is an integrable quench?},
Nucl. Phys. B \textbf{925}, 362-402 (2017)
%doi:10.1016/j.nuclphysb.2017.10.012
[arXiv:1709.04796].

\bibitem{Pozsgay:2018dzs}
B.~Pozsgay, L.~Piroli and E.~Vernier,
\textit{Integrable Matrix Product States from boundary integrability},
SciPost Phys. \textbf{6}, no.5, 062 (2019)
%doi:10.21468/SciPostPhys.6.5.062
[arXiv:1812.11094].

\bibitem{Caux:2013ra}
J.-S.~Caux and F.~H.~L.~Essler,
\textit{Time evolution of local observables after quenching to an integrable model},
Phys. Rev. Lett. \textbf{110} (2013) no.25, 257203
%doi:10.1103/PhysRevLett.110.257203
[arXiv:1301.3806].

\bibitem{Wouters:2014}
B.~Wouters, J.~De Nardis, M.~Brockmann, D.~Fioretto, M.~Rigol and J.-S.~Caux, 
\textit{Quenching the anisotropic Heisenberg chain: exact solution and generalized Gibbs ensemble predictions},
Phys. Rev. Lett. \textbf{113} (2014) 117202 [arXiv:1405.0172].

\bibitem{Pozsgay:2014}
B.~Pozsgay, M.~Mesty{\'a}n, M.~A.~Werner, M.~Kormos, G.~Zar{\'a}nd and G.~Tak{\'a}cs, 
\textit{Correlations after quantum quenches in the XXZ spin chain: failure of the generalized Gibbs ensemble},
Phys. Rev. Lett. \textbf{113} (2014) 117203 [arXiv:1405.2843].

\bibitem{Mestyan:2017xyk}
M.~Mesty{\'a}n, B.~Bertini, L.~Piroli and P.~Calabrese,
\textit{Exact solution for the quench dynamics of a nested integrable system},
J. Stat. Mech. \textbf{1708} (2017) no.8, 083103
%doi:10.1088/1742-5468/aa7df0
[arXiv:1705.00851].

\bibitem{Piroli:2018ksf}
L.~Piroli, E.~Vernier, P.~Calabrese and B.~Pozsgay,
\textit{Integrable quenches in nested spin chains I: the exact steady states},
J. Stat. Mech. \textbf{1906} (2019) no.6, 063103
%doi:10.1088/1742-5468/ab1c51
[arXiv:1811.00432].

\bibitem{Piroli:2018don}
L.~Piroli, E.~Vernier, P.~Calabrese and B.~Pozsgay,
\textit{Integrable quenches in nested spin chains II: fusion of boundary transfer matrices},
J. Stat. Mech. \textbf{1906} (2019) no.6, 063104
%doi:10.1088/1742-5468/ab1c52
[arXiv:1812.05330].

\bibitem{Rylands:2022gev}
C.~Rylands, B.~Bertini and P.~Calabrese,
\textit{Integrable quenches in the Hubbard model},
J. Stat. Mech. \textbf{2210} (2022), 103103
%doi:10.1088/1742-5468/ac98be
[arXiv:2206.07985].

\bibitem{Rylands:2022naf}
C.~Rylands, P.~Calabrese and B.~Bertini,
\textit{Solution of the BEC to BCS Quench in One Dimension},
Phys. Rev. Lett. \textbf{130} (2023) no.2, 023001
%doi:10.1103/PhysRevLett.130.023001
[arXiv:2209.00956 ].

\bibitem{deLeeuw:2015hxa}  
M.~de Leeuw, C.~Kristjansen and K.~Zarembo,
\textit{One-point Functions in Defect CFT and Integrability},
JHEP \textbf{08} (2015), 098
%doi:10.1007/JHEP08(2015)098
[arXiv:1506.06958].

\bibitem{Buhl-Mortensen:2015gfd}
I.~Buhl-Mortensen, M.~de Leeuw, C.~Kristjansen and K.~Zarembo,
\textit{One-point Functions in AdS/dCFT from Matrix Product States},
JHEP \textbf{02} (2016), 052
%doi:10.1007/JHEP02(2016)052
[arXiv:1512.02532].

\bibitem{deLeeuw:2016umh}
M.~de Leeuw, C.~Kristjansen and S.~Mori,
\textit{AdS/dCFT one-point functions of the SU(3) sector},
Phys. Lett. B \textbf{763} (2016), 197-202
%doi:10.1016/j.physletb.2016.10.044
[arXiv:1607.03123].

\bibitem{deLeeuw:2017dkd}
M.~de Leeuw, A.~C.~Ipsen, C.~Kristjansen, K.~E.~Vardinghus and M.~Wilhelm,
\textit{Two-point functions in AdS/dCFT and the boundary conformal bootstrap equations},
JHEP \textbf{08} (2017), 020
%doi:10.1007/JHEP08(2017)020
[arXiv:1705.03898].

\bibitem{Buhl-Mortensen:2017ind}
I.~Buhl-Mortensen, M.~de Leeuw, A.~C.~Ipsen, C.~Kristjansen and M.~Wilhelm,
\textit{Asymptotic One-Point Functions in Gauge-String Duality with Defects},
Phys. Rev. Lett. \textbf{119} (2017) no.26, 261604
%doi:10.1103/PhysRevLett.119.261604
[arXiv:1704.07386].

\bibitem{DeLeeuw:2018cal}
M.~De Leeuw, C.~Kristjansen and G.~Linardopoulos,
\textit{Scalar one-point functions and matrix product states of AdS/dCFT},
Phys. Lett. B \textbf{781} (2018), 238-243
%doi:10.1016/j.physletb.2018.03.083
[arXiv:1802.01598].

\bibitem{Komatsu:2020sup}
S.~Komatsu and Y.~Wang,
\textit{Non-perturbative defect one-point functions in planar $\mathcal{N}=4$  super-Yang-Mills},
Nucl. Phys. B \textbf{958} (2020), 115120
%doi:10.1016/j.nuclphysb.2020.115120
[arXiv:2004.09514 [hep-th]].

\bibitem{Kristjansen:2020mhn}
C.~Kristjansen, D.~M{\"u}ller and K.~Zarembo,
\textit{Integrable boundary states in D3-D5 dCFT: beyond scalars},
JHEP \textbf{08} (2020), 103
%doi:10.1007/JHEP08(2020)103
[arXiv:2005.01392].

\bibitem{Holguin:2025bfe}
A.~Holguin and H.~Kawai,
\textit{Integrability and conformal blocks for surface defects in $\mathcal{N}=4$ SYM},
JHEP \textbf{11} (2025), 043
%doi:10.1007/JHEP11(2025)043
[arXiv:2503.09944].

\bibitem{Chalabi:2025nbg}       
A.~Chalabi, C.~Kristjansen and C.~Su,
\textit{Integrable corners in the space of Gukov-Witten surface defects},
Phys. Lett. B \textbf{866} (2025), 139512
%doi:10.1016/j.physletb.2025.139512
[arXiv:2503.22598].

\bibitem{Linardopoulos:2021rfq}
G.~Linardopoulos and K.~Zarembo,
\textit{String integrability of defect CFT and dynamical reflection matrices},
JHEP \textbf{05} (2021), 203
%doi:10.1007/JHEP05(2021)203
[arXiv:2102.12381].

\bibitem{Linardopoulos:2025ypq}
G.~Linardopoulos,
\textit{String theory methods for defect CFTs},
[arXiv:2501.11985].

\bibitem{Jiang:2019xdz}
Y.~Jiang, S.~Komatsu and E.~Vescovi,
\textit{Structure constants in $ \mathcal{N} $ = 4 SYM at finite coupling as worldsheet g-function},
JHEP \textbf{07} (2020) no.07, 037
%doi:10.1007/JHEP07(2020)037
[arXiv:1906.07733].

\bibitem{Jiang:2019zig}
Y.~Jiang, S.~Komatsu and E.~Vescovi,
\textit{Exact Three-Point Functions of Determinant Operators in Planar $N=4$ Supersymmetric Yang-Mills Theory},
Phys. Rev. Lett. \textbf{123} (2019) no.19, 191601
%doi:10.1103/PhysRevLett.123.191601
[arXiv:1907.11242].

\bibitem{DeLeeuw:2019ohp}
M.~De Leeuw, T.~Gombor, C.~Kristjansen, G.~Linardopoulos and B.~Pozsgay,
\textit{Spin Chain Overlaps and the Twisted Yangian},
JHEP \textbf{01} (2020), 176
%doi:10.1007/JHEP01(2020)176
[arXiv:1912.09338].

\bibitem{Kristjansen:2023ysz}
C.~Kristjansen and K.~Zarembo,
\textit{{\textquoteright}t Hooft loops and integrability},
JHEP \textbf{08} (2023), 184
%doi:10.1007/JHEP08(2023)184
[arXiv:2305.03649].

\bibitem{Ivanovskiy:2024vel}
V.~Ivanovskiy, S.~Komatsu, V.~Mishnyakov, N.~Terziev, N.~Zaigraev and K.~Zarembo,
\textit{Vacuum Condensates on the Coulomb Branch},
[arXiv:2405.19043].

\bibitem{Gombor:2024api}
T.~Gombor and Z.~Bajnok,
\textit{Dual overlaps and finite coupling {\textquoteright}t Hooft loops},
JHEP \textbf{12} (2024), 034
%doi:10.1007/JHEP12(2024)034
[arXiv:2408.14901].

\bibitem{Kristjansen:2024map}
C.~Kristjansen and K.~Zarembo,
\textit{{\textquoteright}t Hooft loops in N=4 super-Yang-Mills},
JHEP \textbf{02} (2025), 179
%doi:10.1007/JHEP02(2025)179
[arXiv:2412.01972].

\bibitem{Coronado:2025xwk}
F.~Coronado, S.~Komatsu and K.~Zarembo,
\textit{Coulomb branch and integrability},
JHEP \textbf{10} (2025), 143
%doi:10.1007/JHEP10(2025)143
[arXiv:2506.07222].

\bibitem{Demjaha:2025axy}
R.~Demjaha and K.~Zarembo,
\textit{String integrability on the Coulomb branch},
JHEP \textbf{09} (2025), 154
%doi:10.1007/JHEP09(2025)154
[arXiv:2506.17955].

\bibitem{Gombor:2025qvk}
T.~Gombor and A.~Holguin,
\textit{Boundary integrability from the fuzzy three sphere},
Phys. Lett. B \textbf{872}, 140078 (2026)
%doi:10.1016/j.physletb.2025.140078
[arXiv:2510.27463].

\bibitem{Gombor:2021uxz}
T.~Gombor and B.~Pozsgay,
\textit{On factorized overlaps: Algebraic Bethe Ansatz, twists, and Separation of Variables},
Nucl. Phys. B \textbf{967} (2021), 115390
%doi:10.1016/j.nuclphysb.2021.115390
[arXiv:2101.10354].

\bibitem{Gombor:2021hmj}
T.~Gombor,
\textit{On exact overlaps for gl(N) symmetric spin chains},
Nucl. Phys. B \textbf{983} (2022), 115909
%doi:10.1016/j.nuclphysb.2022.115909
[arXiv:2110.07960].

\bibitem{Gombor:2023bez}
T.~Gombor,
\textit{Exact overlaps for all integrable two-site boundary states of $ \mathfrak{gl} $(N) symmetric spin chains},
JHEP \textbf{05} (2024), 194
%doi:10.1007/JHEP05(2024)194
[arXiv:2311.04870].

\bibitem{Gombor:2024iix}
T.~Gombor,
\textit{Exact Overlaps for All Integrable Matrix Product States of Rational Spin Chains},
Phys. Rev. Lett. \textbf{135} (2025) no.15, 150402
%doi:10.1103/4vy2-8cnk
[arXiv:2410.23282].

\bibitem{Gombor:2025wvu}
T.~Gombor,
\textit{Derivations for the MPS overlap formulas of rational spin chains},
JHEP \textbf{10} (2025), 035
%doi:10.1007/JHEP10(2025)035
[arXiv:2505.20234].

\bibitem{Aharony:2008ug}
O.~Aharony, O.~Bergman, D.~L.~Jafferis and J.~Maldacena,
\textit{N=6 superconformal Chern-Simons-matter theories, M2-branes and their gravity duals},
JHEP \textbf{10}, 091 (2008)
%doi:10.1088/1126-6708/2008/10/091
[arXiv:0806.1218 [hep-th]].

\bibitem{Minahan:2008hf}
J.~A.~Minahan and K.~Zarembo,
\textit{The Bethe ansatz for superconformal Chern-Simons},
JHEP \textbf{09}, 040 (2008)
%doi:10.1088/1126-6708/2008/09/040
[arXiv:0806.3951 [hep-th]].

\bibitem{Bak:2008cp}
D.~Bak and S.~J.~Rey,
\textit{Integrable Spin Chain in Superconformal Chern-Simons Theory},
JHEP \textbf{10}, 053 (2008)
%doi:10.1088/1126-6708/2008/10/053
[arXiv:0807.2063 [hep-th]].

\bibitem{Yang:2021hrl}
P.~Yang, Y.~Jiang, S.~Komatsu and J.~B.~Wu,
\textit{Three-point functions in ABJM and Bethe Ansatz},
JHEP \textbf{01} (2022), 002
%doi:10.1007/JHEP01(2022)002
[arXiv:2103.15840].

\bibitem{Kristjansen:2021abc}
C.~Kristjansen, D.~L.~Vu and K.~Zarembo,
\textit{Integrable domain walls in ABJM theory},
JHEP \textbf{02} (2022), 070
%doi:10.1007/JHEP02(2022)070
[arXiv:2112.10438].

\bibitem{Gombor:2022aqj}   
T.~Gombor and C.~Kristjansen,
\textit{Overlaps for matrix product states of arbitrary bond dimension in ABJM theory},
Phys. Lett. B \textbf{834} (2022), 137428
%doi:10.1016/j.physletb.2022.137428
[arXiv:2207.06866].

\bibitem{Yang:2022dlk}
P.~Yang,
\textit{Integrable boundary states from maximal giant gravitons in ABJM theory},
Phys. Lett. B \textbf{846} (2023), 138194
%doi:10.1016/j.physletb.2023.138194
[arXiv:2208.12010].

\bibitem{Jiang:2023cdm}
Y.~Jiang, J.~B.~Wu and P.~Yang,
\textit{Wilson-loop one-point functions in ABJM theory},
JHEP \textbf{09} (2023), 047
%doi:10.1007/JHEP09(2023)047
[arXiv:2306.05773].

\bibitem{Wu:2024uix}
J.~B.~Wu and P.~Yang,
\textit{Three-point functions in Aharony-Bergman-Jafferis-Maldacena theory and integrable boundary states},
JHEP \textbf{02} (2025), 030
%doi:10.1007/JHEP02(2025)030
[arXiv:2408.03643].

\bibitem{Bai:2024qtg}
N.~Bai and M.~Z.~Shao,
\textit{Integrable matrix product states of ABJM theory from projecting method},
Mod. Phys. Lett. A \textbf{40} (2025) no.19n20, 2550068
%doi:10.1142/S0217732325500683
[arXiv:2411.09282].

\bibitem{Liu:2025uiu}
Y.~Liu and J.~B.~Wu,
\textit{Chiral Integrable Boundary States in the SU(4) Alternating Spin Chain},
Sci. China Phys. Mech. Astron. \textbf{69}, 231011 (2026)
%doi.org/10.1007/s11433-025-2834-3
[arXiv:2507.03489].

\bibitem{Liu:2026abjmre}
Y.~Liu, N.~Bai, M.~Z.~Shao and J.~B.~Wu,
\textit{Chiral Integrable Boundary States of ABJM Spin Chain from Reflection Equations},
[arXiv:2602.01697].

\bibitem{Linardopoulos:2022wol}  
G.~Linardopoulos,
\textit{String integrability of the ABJM defect},
JHEP \textbf{06} (2022), 033
%doi:10.1007/JHEP06(2022)033
[arXiv:2202.06824].


\bibitem{Yang:2021kot}
P.~Yang, Y.~Jiang, S.~Komatsu and J.~B.~Wu,
\textit{D-branes and orbit average},
SciPost Phys. \textbf{12}, no.2, 055 (2022)
%doi:10.21468/SciPostPhys.12.2.055
[arXiv:2103.16580].

\bibitem{Gombor:2020kgu}
T.~Gombor and Z.~Bajnok,
\textit{Boundary states, overlaps, nesting and bootstrapping AdS/dCFT},
JHEP \textbf{10} (2020), 123
%doi:10.1007/JHEP10(2020)123
[arXiv:2004.11329].


\bibitem{Mezincescu:1991ke}
L.~Mezincescu and R.~I.~Nepomechie,
\textit{Fusion procedure for open chains},
J. Phys. A \textbf{25} (1992), 2533-2544

\bibitem{Bai:2025mpw}
N.~Bai,
\textit{A general fusion procedure for open $\mathfrak{g}\mathfrak{l}(N)$ spin chains: application to the ABJM spin chain},
JHEP \textbf{01} (2026), 075
%doi:10.1007/JHEP01(2026)075
[arXiv:2507.19394].

\bibitem{Arnaudon:2004sd}
D.~Arnaudon, J.~Avan, N.~Crampe, A.~Doikou, L.~Frappat and E.~Ragoucy,
\textit{General boundary conditions for the sl(N) and sl(M|N) open spin chains},
J. Stat. Mech. \textbf{0408} (2004), P08005
%doi:10.1088/1742-5468/2004/08/P08005
[arXiv:math-ph/0406021].

\bibitem{Molev 2002}
A.~I.~Molev and E.~Ragoucy, \textit{Representations of reflection algebras}, Rev. Math. Phys. \textbf{14} (2002) 317 [arXiv:math/0107213]

\end{thebibliography}
\end{document}